\documentclass[12pt,preprint]{aastex}

\usepackage{graphicx,psfig,amssymb,verbatim,natbib}

\newcommand{\lya}{Ly$\alpha$}

\newcommand\gp{$g_{475}$}

\newcommand\ip{$i_{775}$}
\newcommand\inp{$I_{814}$}
\newcommand\zp{$z_{850}$}
\newcommand\vp{$V_{606}$}
\newcommand\bp{$B_{435}$}
\newcommand\civ{\hbox{C$\rm IV~\lambda1549$\AA}}
\newcommand\heii{\hbox{He$\rm II~\lambda1640$\AA}}
\newcommand\se{Section~}
\newcommand\hi{\hbox{H$\rm I$}}
\newcommand\hii{\hbox{H$\rm II$}}

\shorttitle{Clustering of Star-forming Galaxies Near a Radio Galaxy at $z=5.2$}
\shortauthors{Overzier et al.}

\begin{document}


\title{Clustering of Star-forming Galaxies Near a Radio Galaxy at $z=5.2$\altaffilmark{1}}


\author{Roderik A. Overzier\altaffilmark{2},
G.K. Miley\altaffilmark{2},
R.J. Bouwens\altaffilmark{3},
N.J.G. Cross\altaffilmark{4},
A.W. Zirm\altaffilmark{2},
N. Ben\'{\i}tez\altaffilmark{5},
J.P. Blakeslee\altaffilmark{6},
M. Clampin\altaffilmark{7},
R. Demarco\altaffilmark{6},
H.C. Ford\altaffilmark{6},
G.F. Hartig\altaffilmark{8},
G.D. Illingworth\altaffilmark{3},
A.R. Martel\altaffilmark{6},
H.J.A. R\"{o}ttgering\altaffilmark{2},
B. Venemans\altaffilmark{2},
D.R. Ardila\altaffilmark{6},
F. Bartko\altaffilmark{9}, 
L.D. Bradley\altaffilmark{6},
T.J. Broadhurst\altaffilmark{10},
D. Coe\altaffilmark{6},
P.D. Feldman\altaffilmark{6},
M. Franx\altaffilmark{2},
D.A. Golimowski\altaffilmark{6},
T. Goto\altaffilmark{6},
C. Gronwall\altaffilmark{11},
B. Holden\altaffilmark{3},
N. Homeier\altaffilmark{6},
L. Infante\altaffilmark{12}
R.A. Kimble\altaffilmark{7},
J.E. Krist\altaffilmark{13},
S. Mei\altaffilmark{6},
F. Menanteau\altaffilmark{6},
G.R. Meurer\altaffilmark{6},
V. Motta\altaffilmark{6,12},
M. Postman\altaffilmark{8},
P. Rosati\altaffilmark{14}, 
M. Sirianni\altaffilmark{6}, 
W.B. Sparks\altaffilmark{8}, 
H.D. Tran\altaffilmark{15}, 
Z.I. Tsvetanov\altaffilmark{6},   
R.L. White\altaffilmark{8}
\& W. Zheng\altaffilmark{6}}


\altaffiltext{1}{Based on observations made with the NASA/ESA Hubble Space Telescope, which is operated by the Association of Universities for Research in Astronomy, Inc., under NASA contract NAS 5-26555. These observations are associated with program \# 9291}

\altaffiltext{2}{Leiden Observatory, Postbus 9513, 2300 RA Leiden, Netherlands.}
\email{overzier@strw.leidenuniv.nl}

\altaffiltext{3}{UCO/Lick Observatory, University of California, Santa
Cruz, CA 95064.}

\altaffiltext{4}{Royal Observatory Edinburgh, Blackford Hill, Edinburgh, EH9 3HJ, UK}

\altaffiltext{5}{Inst. Astrof\'{\i}sica de Andaluc\'{\i}a (CSIC), Camino Bajo de Hu\'{e}tor, 24, Granada 18008, Spain}

\altaffiltext{6}{Department of Physics and Astronomy, The Johns Hopkins
University, 3400 North Charles Street, Baltimore, MD 21218.}

\altaffiltext{7}{NASA Goddard Space Flight Center, Code 681, Greenbelt, MD 20771.}

\altaffiltext{8}{STScI, 3700 San Martin Drive, Baltimore, MD 21218.}

\altaffiltext{9}{Bartko Science \& Technology, 14520 Akron Street, 
Brighton, CO 80602.}    

\altaffiltext{10}{Racah Institute of Physics, The Hebrew University,
Jerusalem, Israel 91904.}

\altaffiltext{11}{Department of Astronomy and Astrophysics, The
Pennsylvania State University, 525 Davey Lab, University Park, PA
16802.}


\altaffiltext{12}{Departmento de Astronom\'{\i}a y Astrof\'{\i}sica,
Pontificia Universidad Cat\'{o}lica de Chile, Casilla 306, Santiago
22, Chile.}

\altaffiltext{13}{Jet Propulsion Laboratory, M/S 183-900, 4800 Oak Grove Drive, Pasadena, CA 91109}

\altaffiltext{14}{European Southern Observatory,
Karl-Schwarzschild-Strasse 2, D-85748 Garching, Germany.}


\altaffiltext{15}{W. M. Keck Observatory, 65-1120 Mamalahoa Hwy., Kamuela, HI 96743}







\begin{abstract}
We present HST/ACS observations of the most distant radio galaxy known, 
TN J0924--2201 at $z=5.2$. This radio galaxy has 6 spectroscopically confirmed \lya\ 
emitting companion galaxies, and appears to lie within an overdense region. The radio galaxy 
is marginally resolved in \ip\ and \zp\ showing continuum emission aligned 
with the radio axis, similar to what is observed for lower redshift radio galaxies. 
Both the half-light radius and the UV star formation rate are comparable to the typical 
values found for Lyman break galaxies at $z\sim4-5$. 
The \lya\ emitters are sub-$L^*$ galaxies, 
with deduced star formation rates of $1-10$ $M_\odot$ yr$^{-1}$.  
One of the \lya\ emitters is only detected in \lya. Based on the star 
formation rate of $\sim3$ M$_\odot$ yr$^{-1}$ calculated from \lya, the lack of continuum emission 
could be explained if the galaxy is younger than $\sim2$ Myr and is producing its 
first stars.

Observations in \vp\ip\zp\ were used to identify additional Lyman break galaxies associated 
with this structure. In addition to the radio galaxy, there are 22 \vp-break ($z\sim5$) 
galaxies with \zp$<$26.5 (5$\sigma$), two of which are also in the spectroscopic sample.
We compare the surface density of $\sim2$ arcmin$^{-2}$ to that of   
similarly selected \vp-dropouts extracted from GOODS and the UDF Parallel fields. We find  
evidence for an overdensity to very high confidence ($>99$\%), based on a counts-in-cells analysis 
applied to the control field. The excess is suggestive of the \vp-break objects being associated with 
a forming cluster around the radio galaxy.

\end{abstract}





\keywords{cosmology: observations -- early universe -- large-scale structure of universe -- galaxies: high-redshift -- galaxies: clusters: general -- galaxies: starburst -- galaxies: individual (TN J0924--2201)}


\section{Introduction}

Where can we find the progenitors of the galaxy clusters that populate the local Universe? 
The evolution of rich galaxy clusters has been studied out to $z\sim1.4$ \citep{mullis05}. These clusters 
have been discovered primarily via their bright X-ray emission, the signature of 
virialised gas within a deep gravitational potential well. Follow-up 
observations have revealed that some of the galaxy populations in distant clusters are 
relatively old, as evidenced by, for example, the tight scatter in the color-magnitude relation 
for early-type galaxies \citep[e.g.][]{stanford98,blakeslee03a,wuyts04,holden05}, and the mild 
evolution of the morphology density relation for cluster ellipticals since $z\sim1$ \citep{postman04}. 
This suggests that an interesting epoch of cluster formation could lie at higher redshifts.  
Several good examples for overdensities of galaxies at $1.5\lesssim z\lesssim6$, possibly the progenitors of clusters, 
exist in the literature \citep[e.g.][]{pascarelle96,steidel98,keel99,steidel00,francis01,moller01,sanchez02,shimasaku03,ouchi05,steidel05}. These structures have been found often as by-products of large-area field surveys using broad or narrow band imaging, or by targeting luminous radio sources.

One technique for finding distant galaxy overdensities is based on the empirical evidence 
that powerful radio galaxies are among the most massive forming galaxies at high redshift 
\citep[e.g.][]{debreuck02,dey97,pentericci01,zirm03}. In the 
standard cold dark matter (CDM) universe model, massive galaxies and galaxy clusters 
are associated with the most massive dark matter haloes within the large-scale structure. It has been found that massive black 
holes are a key-ingredient of local massive galaxies, and that their mass scales  
in proportion to the mass of the spheroidal component of the host galaxy \citep{magorrian98,gebhardt00,ferraresemerritt00}. 
Radio galaxies may therefore demarcate the location of forming clusters, analogous to the suggested scaling relations between halo, host 
galaxy and black hole mass at low redshift. A program with the Very Large Telescope (VLT) of the 
European Southern Observatory to search for galaxy overdensities around luminous high-redshift 
radio galaxies through deep narrow band \lya\ imaging and spectroscopy has indeed revealed 
that the radio galaxies are often accompanied by large numbers of line emitting galaxies 
\citep{pentericci00,kurk03,venemans02,venemans04,venemans05}.

We have started a study with the {\it Advanced Camera for Surveys} on the Hubble Space Telescope \citep[HST/ACS;][]{ford98} to survey 
some of these \lya-selected {\it protoclusters}\footnote{The term {\it protocluster} has no strict definition in 
literature. It is commonly used to describe galaxy overdensities at high redshift with mass estimates that are comparable to those of 
galaxy clusters, but without any detectable X-ray emission from a hot, virialised intra-cluster medium.}. 
Our goal is to augment our study of emission line 
objects by deep broad band observations to search for Lyman break galaxies (LBGs). 
Observations of the radio 
galaxy protocluster TN J1338--1942 at $z=4.1$ have shown that the overdensity of \lya\ emitters 
discovered by \citet{venemans02} is accompanied by a similar overdensity of Lyman break galaxies, 
allowing us to assess distinct galaxy populations in overdense regions \citep[][Overzier et al., in prep.]{miley04}. 
The radio galaxy TN J1338--1942 was found to have a complex morphology, showing clear signs of AGN 
feedback on the forming ISM and a starburst-driven wind possibly feeding the gaseous halo that 
surrounds the galaxy \citep{zirm05}. The \lya\ emitters have relatively faint UV continua 
and small angular sizes compared to the generally brighter LBG population in field studies \citep[e.g.][]{ferguson04,bouwens04_sizes}.

TN J0924--2201 at $z=5.19$ is the most distant radio galaxy known \citep{vanbreugel99,debreuck00}. 
Following the successes obtained in identifying \lya\ galaxy overdensities 
around our sample of powerful high redshift radio galaxies at $2.2<z<4.1$, \citet{venemans04} have 
probed the distribution of \lya\ emitters around TN J0924--2201: there are 6 spectroscopically 
confirmed companions within a (projected) radius of 2.5 Mpc and a (rest-frame) 1000 km s$^{-1}$ from the 
radio galaxy, corresponding to a surface overdensity of 1.5--6 with respect to the field.
This overdensity is comparable to that of radio galaxy/\lya\ protoclusters at lower $z$, and is in 
support of the idea that radio galaxies conspicuously identify groups or cluster-like regions in the very early Universe. 

In this paper we present the results of a follow-up study of the galaxy overdensity near TN J0924--2201 
through high resolution imaging observations obtained with HST/ACS. The 
primary goal of these observations was to look for an enhancement in the surface density of Lyman break galaxies in the field of  
TN J0924--2201, which would generally be missed by the selection based on the presence of a \lya\ emission line alone. 
LBGs and \lya\ emitters are strongly clustered at $z=3-5$, and are highly biased relative to predictions for the dark matter distribution 
\citep{giavalisco98,adelberger98,ouchi04_r0}. The biasing becomes stronger for galaxies with higher rest-frame UV luminosity \citep{giavalisco01}. 
In an excellent, all-encompassing census of the clustering properties of LBGs, \citet{ouchi04_r0} found that the bias may also increase with redshift 
and dust extinction, in addition to UV luminosity. By comparing the number densities of LBGs to that of dark halos predicted by 
\citet{sheth99} they concluded that $z=4$ LBGs could be hosted by halos of $1\times10^{11}-5\times10^{12}$ M$_\odot$ \citep[see also][]{hamana04}, and that the descendants of those halos at $z=0$ have masses that are comparable to the masses of groups and clusters.   

The structure of this paper is as follows. In \se2 we describe our observations and data analysis. 
\se3 subsequently deals with the host galaxy of the radio source, the spectroscopically 
confirmed \lya\ emitters, and our sample of $z\sim5$ Lyman break galaxies. In \se4 we discuss the evidence that suggests 
that TN J0924--2201 may pinpoint a young galaxy cluster, and we present our conclusions in \se5.
We use a cosmology in which $H_0=72$ km s$^{-1}$ Mpc$^{-1}$, $\Omega_M=0.27$, and $\Omega_\Lambda=0.73$ \citep{spergel03}. 
In this Universe, the luminosity distance is 49.2 Gpc and the angular scale size is 6.2 kpc arcsec$^{-1}$ at $z=5.2$. 
The lookback time is 12.2 Gyr, corresponding to an epoch when the Universe was approximately 8\% of its current age. 

\begin{figure*}[t]
\begin{center}
\includegraphics[width=0.7\textwidth]{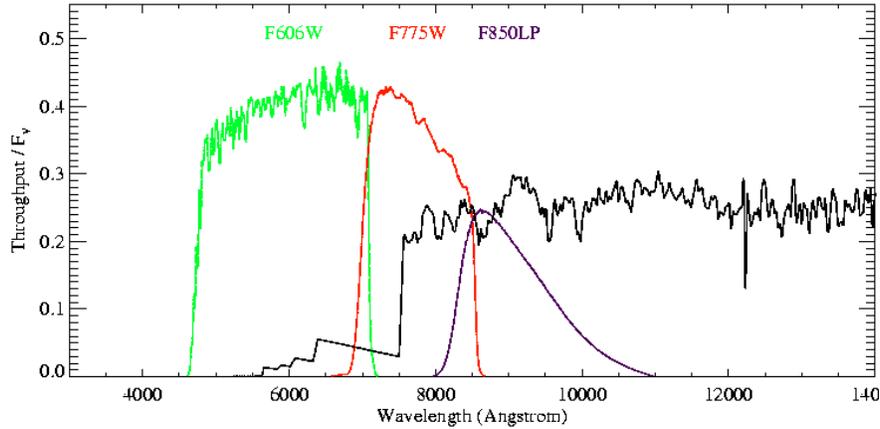}
\end{center}
\caption{\label{fig:filters}Total effective throughput of the HST/ACS filterset used in 
our observations. The SED template shown is the $SB2$ template from \citet{benitez00}, redshifted to 
$z=5.19$ taking into account the attenuation of the IGM following the prescription of \citet{madau96}. 
The Lyman break occurs between filters \vp\ and \ip\ for galaxies at $z\sim5$.
}
\end{figure*}

\section{ACS Observations, data reduction and photometry}

\subsection{TN J0924--2201}

\begin{figure*}[t]
\begin{center}
\includegraphics[width=0.7\textwidth]{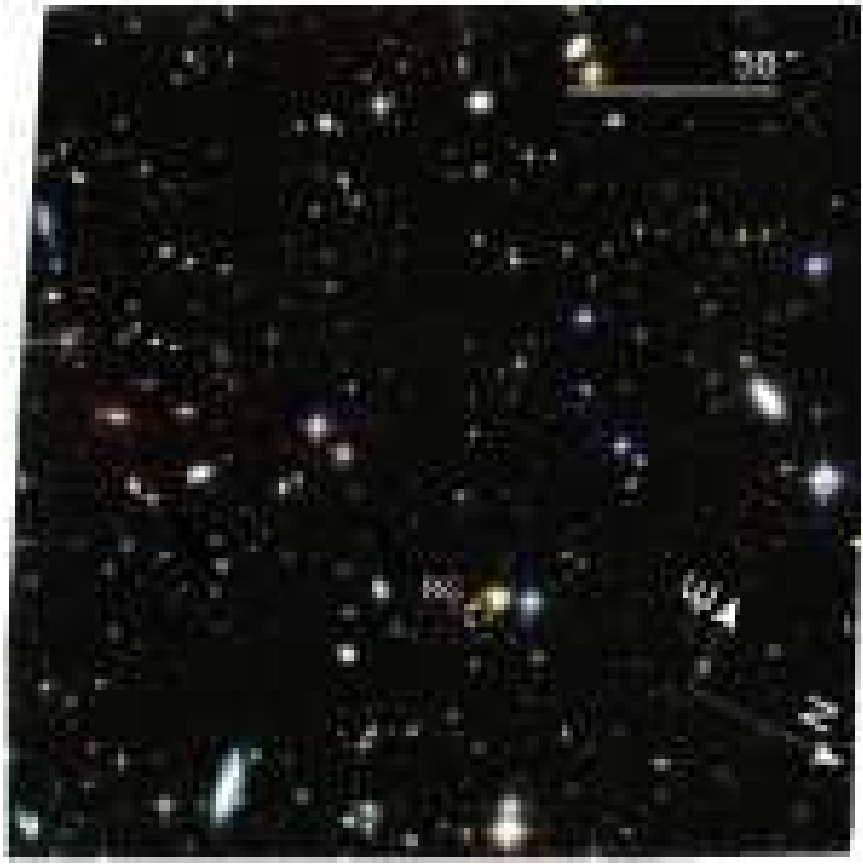}
\end{center}
\caption{\label{fig:colorfield}ACS color image showing \vp\ in blue, \ip\ in green and \zp\ in red. The field measures $11.7$ arcmin$^{2}$. The approximate position of the radio galaxy is marked `RG'.}
\end{figure*}

We surveyed the field surrounding TN J0924--2201 with a single pointing of the Wide-Field Channel (WFC) 
of HST/ACS.   
The field position was chosen in order to maximize the number of spectroscopically-confirmed 
\lya\ emitters in the field. The radio galaxy is located at 
$\alpha_{J2000}=9^h24^m19.90^s$, $\delta_{J2000}=-22\degr01\arcmin42.0\arcsec$, and 4 of the \lya\ emitters fall within 
the 11.7 arcmin$^{2}$ field. The field has an extinction value $E(B-V)=0.057$ determined from the dust maps of 
\citet{schlegel98}. The observations were carried out between 29 May and 8 June 2003, 
as part of the ACS Guaranteed Time Observing (GTO) high redshift cluster program. The total observing time 
of 14 orbits was split over the \vp\ (9400 s), \ip\ (11800 s) and \zp\ (11800 s) broad-band filters, bracketing 
redshifted Ly$\alpha$ at 7527 \AA. The filter transmission curves are indicated in Fig. \ref{fig:filters}. 
Each orbit was split into two 1200 s exposures to facilitate the removal of cosmic rays. 
A color image of the field is shown in Fig. \ref{fig:colorfield}.

The data were processed through CALACS at STScI and the ACS pipeline science investigation software {\it Apsis}
\citep[][]{blakeslee03b} that was developed by and for the ACS GTO team. By default {\it Apsis} provides final drizzled 
images with a pixel scale of 0\farcs05 pixel$^{-1}$. However, to match the image scale of the public data release of the GOODS 
data (our main comparison dataset), we drizzled the science images onto a frame with a pixel scale of 0\farcs03 pixel$^{-1}$. 
Fig. \ref{fig:maglim} (left panels) shows 
the limiting magnitudes for each filter as a function of aperture diameter and signal-to-noise. The $2\sigma$ limiting 
magnitudes in a square aperture of $0.2$ arcsec$^2$ are $\sim29.0$ in \vp, $\sim28.5$ in \ip, and $\sim28.0$ in \zp. There is no significant difference in the detection limits for the 0\farcs05 pixel$^{-1}$ 
dataset and the 0\farcs03 pixel$^{-1}$ dataset. The total filter exposure times, extinctions and zeropoints \citep{sirianni05} are listed in Table \ref{tab:log2}. 

\subsection{GOODS public data}

We used the public imaging data from the Great Observatories Origins Deep Survey 
\citep[GOODS;][]{giavalisco04_survey} as a control field for our data. Similar to TN J0924--2201, 
GOODS has observations in \vp, \ip\ and \zp, and is comparably deep. We downloaded the V1.0 mosaicked 
images release\footnote{http://www.stsci.edu/science/GOODS/} for the GOODS Chandra Deep Field South (CDF-S) 
and Hubble Deep Field North (HDF-N) regions. Details on how these images were produced can be 
found in \citet{giavalisco04_survey}. In total we used an area of $\sim314$ arcmin$^{2}$ from this survey 
(see \se\ref{sec:overdensity}), roughly $27\times$ larger than a single ACS pointing.  
We used the zeropoints given by \citet{giavalisco04_survey} for this dataset. 
Fig. \ref{fig:maglim} (bottom panels) shows that the depth is comparable to that of our TN J0924--2201 observations. 
A summary of the GOODS observations are provided in Table \ref{tab:log2}. 

\subsection{UDF parallels}

\begin{figure}[t]
\begin{center}
\includegraphics[width=0.5\textwidth]{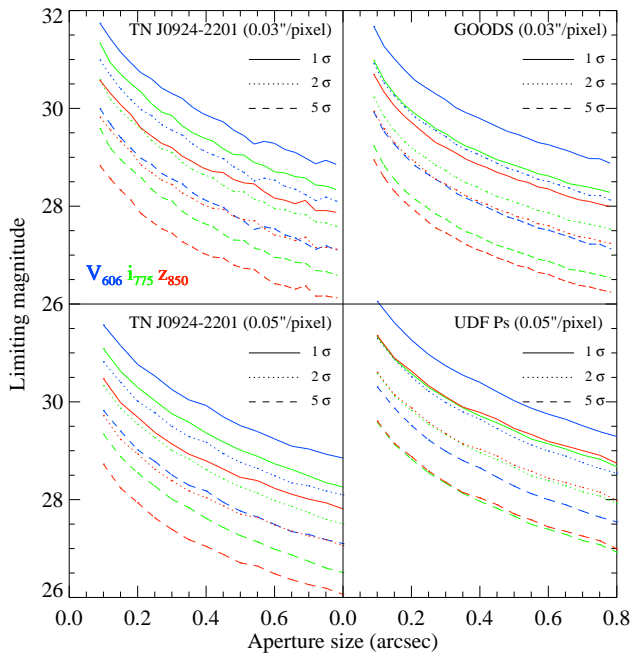}
\end{center}
\caption{\label{fig:maglim}Depth as a function of square aperture diameter for the different datasets. Curves give the $1\sigma$ (solid), 
$2\sigma$ (dotted), and $5\sigma$ (dashed) limiting magnitudes in \vp\ (blue), \ip\ (green), 
and \zp\ (red).}
\end{figure}

We also used the two ACS parallels to the Hubble Ultra Deep Field (UDF) NICMOS observations (GO-9803;
R. I. Thompson et al.) for comparison\footnote{Available through MAST (http://archive.stsci.edu/)}. The observations consist of 2 parallel fields with pointings of 
$\alpha_{J2000}=3^h32^m46.0^s$, $\delta_{J2000}=-27\degr54\arcmin42.3\arcsec$ (UDF-P1) and 
$\alpha_{J2000}=3^h32^m1.0^s$, $\delta_{J2000}=-27\degr48\arcmin3.5\arcsec$ (UDF-P2). The  
heavily dithered images were trimmed down to only the central regions covering $11.7$ arcmin$^{2}$ for each 
parallel field. The data were reduced using {\it Apsis} at the default output scale of 0\farcs05 pixel$^{-1}$.
The UDF parallels reach about 0.5--1 magnitudes deeper in each filter compared to both TN J0924--2201 
and GOODS (see Fig. \ref{fig:maglim}). Details on the observations are given in Table \ref{tab:log2}. 

\subsection{Object detection and photometry}

Object detection and photometry was done using the Source Extractor (SExtractor) software package of \citet{bertin96}. 
We used SExtractor in double-image mode, 
where object detection and aperture determination are carried out on the so-called ``detection image", 
and the photometry is carried out on the individual filter images. The \zp-band was used as the detection image. 
Photometric errors are calculated using the root mean square (RMS) images that contain the final error per 
pixel for each output science image. The RMS images correctly reflect the pixel variation when images are stacked in the absense of 
non-integer pixel shifts or corrections for the geometric distortion. For the fields at a scale of 0\farcs03 pixel$^{-1}$ 
(TN J0924--2201 and GOODS), the main parameters influencing the detection and photometry are essentially the same as the parameters that 
were used to construct the GOODS `r1.1z' public dataset source catalogue \citep[see][]{giavalisco04_survey}: we initially 
considered all detections with a minimum of 16 connected pixels each containing $>$0.6 times the standard deviation of 
the local background (giving a signal-to-noise ratio (S/N) of $>$2.4). SExtractor's deblending 
parameters were set to $\mathtt{DEBLEND\_MINCONT=0.03}$, $\mathtt{DEBLEND\_NTHRESH=32}$. The publicly available 
inverse variance images provided by the GOODS team were converted to RMS images to ensure that the absolute 
standard deviations per pixel are used by SExtractor. For the 
UDF Parallel datasets that were drizzled at a scale of 0\farcs05 pixel$^{-1}$, we 
detected objects using a minimum of 5 connected pixels at a threshold of 
1.1 times the RMS of the local background (nominal S/N of $>$2.4) and setting $\mathtt{DEBLEND\_MINCONT=0.1}$ 
and $\mathtt{DEBLEND\_NTHRESH=8}$. 

After this initial detection we rejected all objects with S/N less than 5 in \zp, where we define S/N as the ratio of 
counts in the isophotal aperture to the errors on the counts.   
The remaining objects were considered real objects. Galactic stars appear to closely overlap with galaxies in the 
\vp--\ip, \ip--\zp\ color-color plane (see section 3.4). We initially rejected all point sources on the basis of high 
SExtractor stellarity index, e.g., setting S/G$<$0.85 (non-stellar objects with high confidence). 

We used SExtractor's $\mathtt{MAG\_AUTO}$ to estimate total object magnitudes within an aperture radius of 
$2.5\times r_{\mathrm{Kron}}$ \citep{kron80}, but calculated {\it galaxy colors} from the isophotal magnitudes 
measured by SExtractor within the aperture defined by the isophotal area of the object in the \zp-band. 
These procedures are optimal for (faint) object detection and aperture photometry with ACS \citep{benitez04}. 
 We measured half-light radii defined as that radius that contains half of the total light using annular photometry 
out to $2.5\times r_{\mathrm{Kron}}$ (performed by SExtractor). The total magnitudes and half-light radii 
have not been corrected for the amount of light missed outside the apertures, unless stated otherwise 
\citep[see][Overzier et al., in prep. for aperture corrections applied to ACS observations]{benitez04,giavalisco04_survey}. 
All colors and magnitudes quoted in this paper have been 
corrected for foreground extinction and are in the $AB$ system of \citet{oke71}.

\begin{figure*}[t]
\begin{center}
\includegraphics[scale=.25]{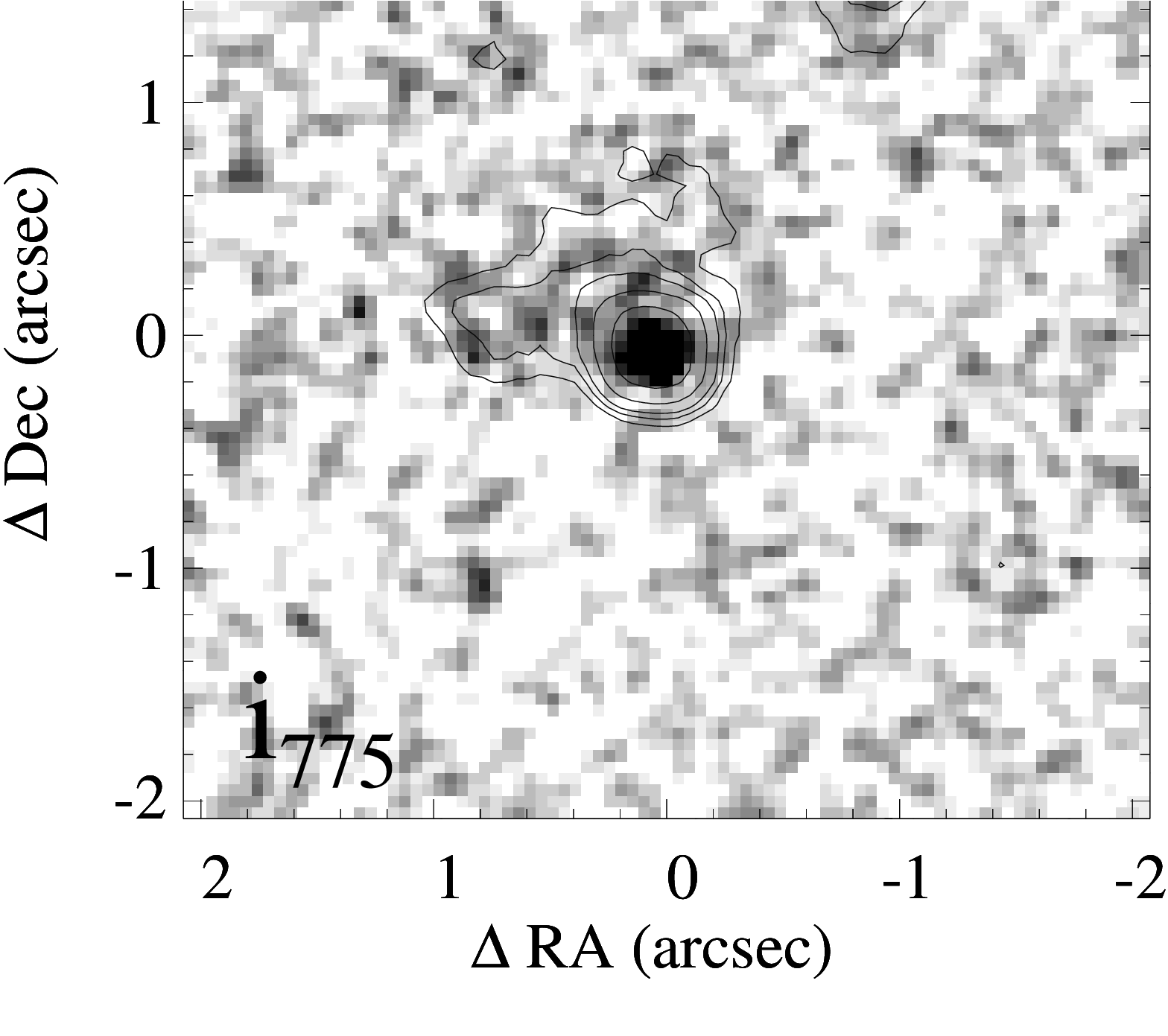}
\includegraphics[scale=.25]{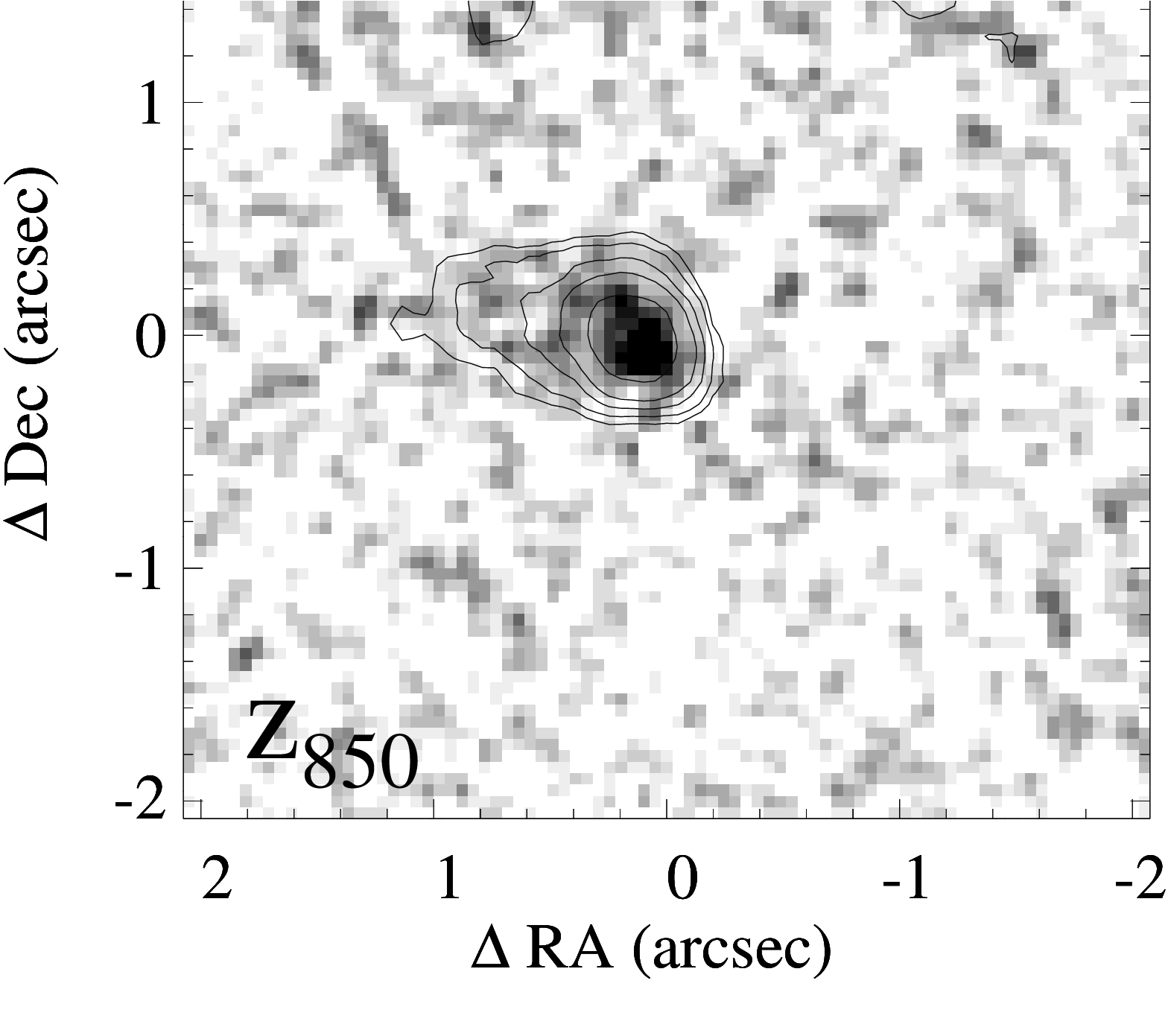}\\
\vspace{10mm}
\includegraphics[scale=.25]{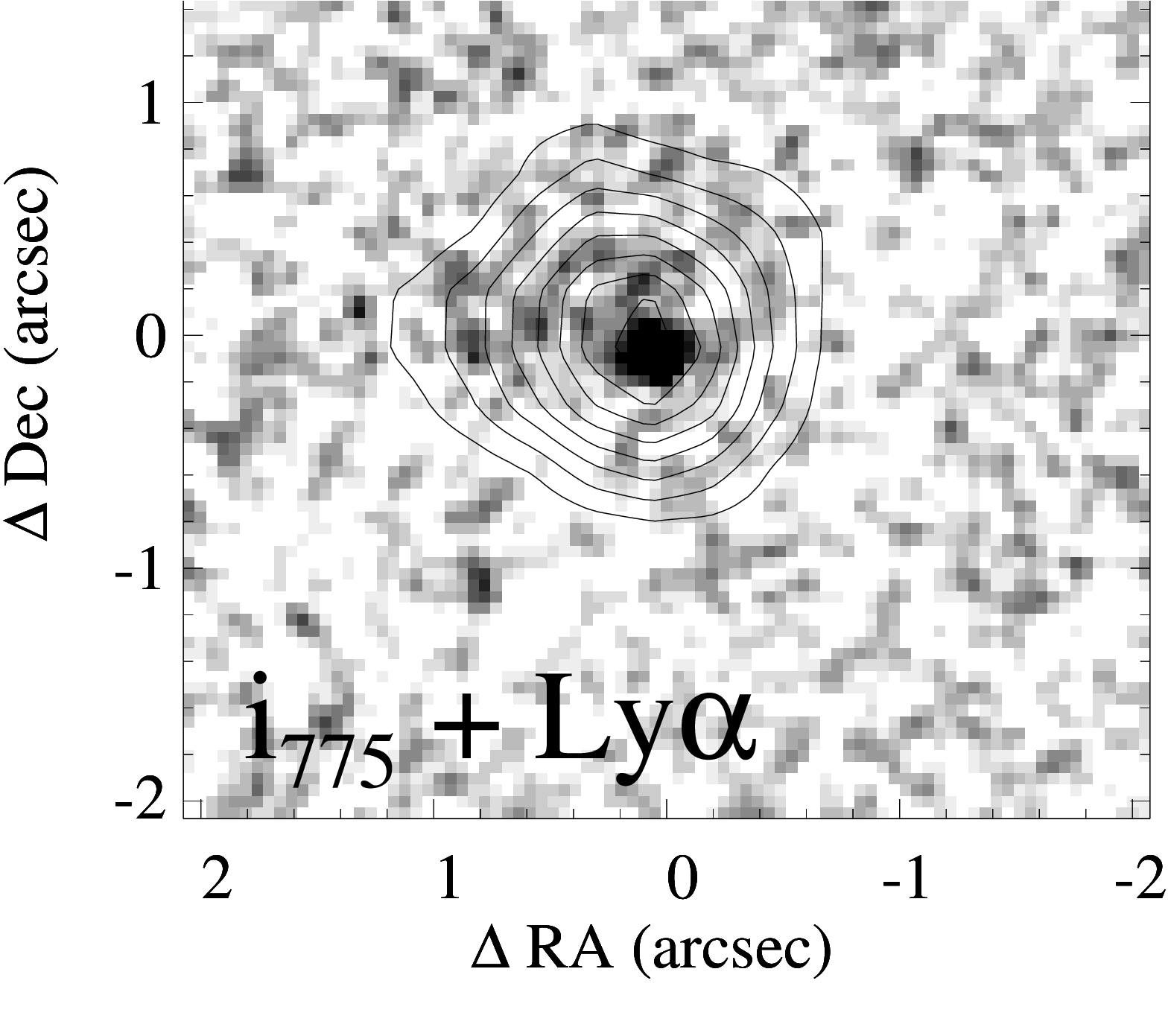}
\includegraphics[scale=.25]{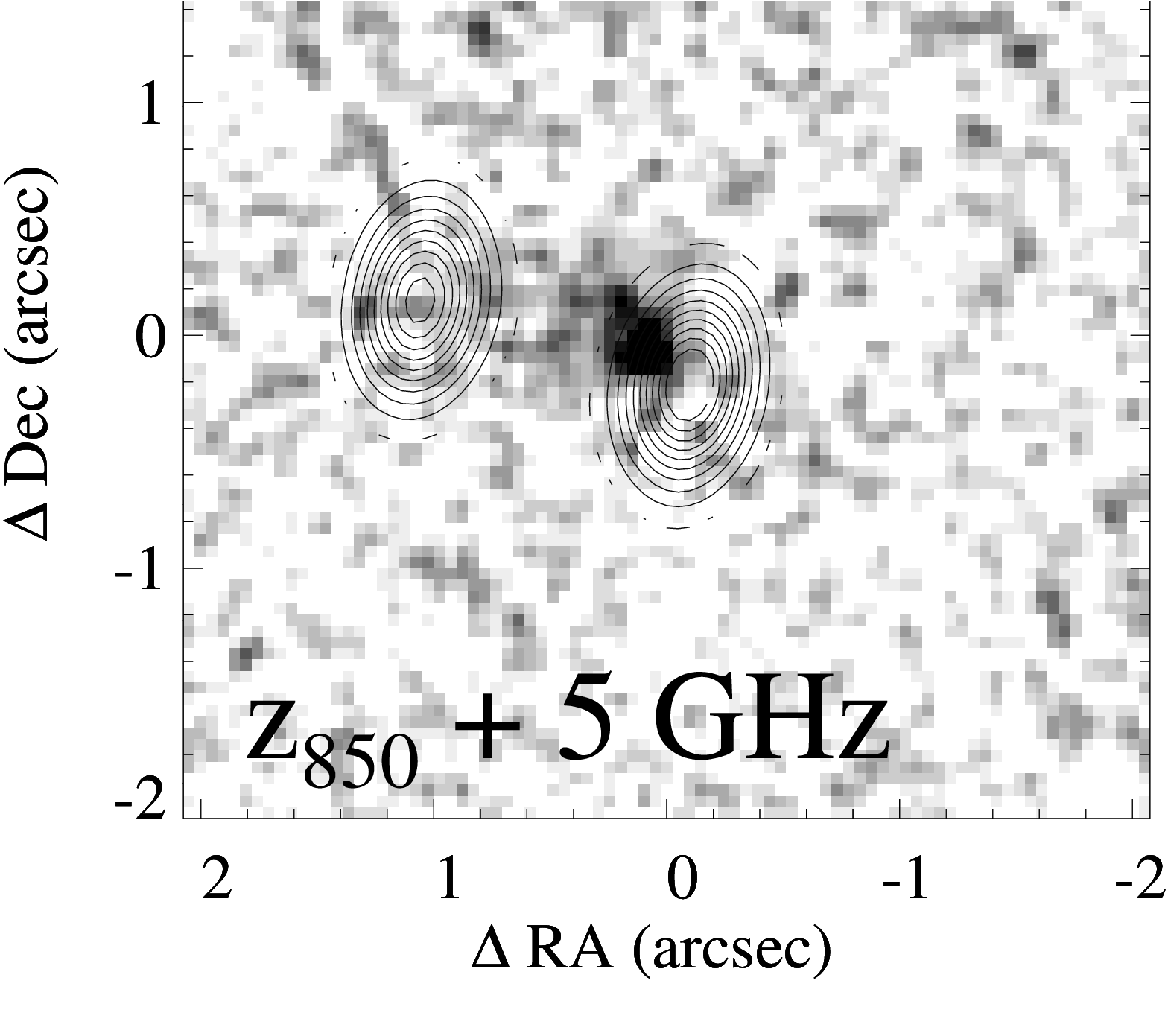}
\end{center}
\caption{\label{fig:rg}HST/ACS images of radio galaxy TN J0924--2201. {\it Top left:} \ip\ with contours of the same image smoothed using a 0.15\arcsec\ (FWHM) Gaussian to enhance the faint `tail'. {\it Top right:} Same as Top left, but for the \zp-band. {\it Bottom left:} \ip-band image in greyscale with contours representing the ground-based narrow-band \lya\ image (with a seeing of $\sim$0.8\arcsec\ (FWHM)). {\it Bottom right} \zp-band image in greyscale with contours of the 4.86 GHz radio image overlayed (C. De Breuck, private communications). Throughout this paper, all ACS geyscale postage stamps 
have been smoothed using a Gaussian kernel of 0.075\arcsec\ (FWHM). The continuum, the \lya, and the radio emission are all aligned.}
\end{figure*}

\subsection{Photometric redshifts}

We used the Bayesian Photometric Redshift code (BPZ) of \citet{benitez00} to obtain 
estimates for galaxy redshifts, $z_B$. For a complete description of BPZ and the robustness of its results, we refer
the reader to \citet{benitez00} and \citet{benitez04}. Our library of galaxy spectra is based on the elliptical, intermediate 
($Sbc$) and late type spiral ($Scd$), and irregular templates of \citet{coleman80}, augmented by starburst galaxy 
templates with $E(B-V)\sim0.3$ ($SB2$) and $E(B-V)\sim0.45$ ($SB3$) from \citet{kinney96}, and two simple stellar 
population (SSP) models with ages of 5 Myr and 25 Myr from \citet{bruzualcharlot03}. The latter two templates have been found to improve 
the accuracy of BPZ for very blue, young high redshift galaxies in the UDF (Coe et al., in prep.). 
BPZ makes use of a parameter `ODDS' defined as 
$P(|z-z_B|<\Delta z)$ that gives the total probability that the true redshift is within an uncertainty $\Delta z$. 
For a Gaussian probability distribution a $2\sigma$ confidence interval centered on $z_B$ would get an ODDS of $>0.95$. 
The empirical accuracy of BPZ is $\sigma\approx0.1(1+z_B)$ for objects with $I_{814}\lesssim24$ and $z\lesssim4$ observed in 
the \bp\vp\inp-bands with ACS to a depth comparable to our observations \citep{benitez04}. 
Note that we will be applying BPZ to generally fainter objects at $z\sim5$ observed in (\bp)\vp\ip\zp. The true accuracy for such a sample has yet to be determined empirically.

\section{Results}

\subsection{Radio galaxy TN J0924--2201}
\label{sec:rg}

The radio galaxy was not detected in \vp\ due to the attenuation of flux shortward of \lya\ 
by the intergalactic medium (IGM). We derive a $2\sigma$ upper limit of 28.3 magnitude within $r_{\mathrm{Kron}}$. 
The galaxy is detected in the other filters with total magnitudes of \ip$=26.0\pm0.1$ and \zp$=25.5\pm0.1$, 
and colors of \vp--\ip$>2.7$ and \ip--\zp$=0.4\pm0.1$ (Table \ref{tab:laes}). 

The radio galaxy's \vp--\ip\ color is 
affected by the relatively large equivalent width of \lya\ ($EW_0=83$ \AA, \citet{venemans04}). 
In the \ip- and \zp-bands, the galaxy consists of a 
compact object with a $\sim1\arcsec$ `tail' extending towards the East that we have made visible by smoothing the ACS images shown in Fig. \ref{fig:rg} using a Gaussian kernel of 0\farcs15 (FWHM). Also shown is the narrow-band \lya\ 
from \citet{venemans04} in contours superposed on the ACS \ip-band (Fig. \ref{fig:rg}, bottom left). 
The narrow-band image was registered to the ACS image using a nearby star $\sim3\arcsec$ to the 
northwest of the radio galaxy. The main component observed with ACS is entirely embedded in the $\sim1\farcs5$ ($\sim9$ kpc) \lya\ 
halo (compared to a seeing of 0.8\arcsec). 
The lower right panel of Fig. \ref{fig:rg} shows the VLA 4.86 GHz radio contours (C. De Breuck, private communications), 
overlayed on the ACS \zp-band image. The relative astrometry 
could not be determined to better than $0\farcs5$. 
We find good correspondence between the orientations of the radio emission and the extended ACS emission. 
This is analogous to the alignment of both the UV continuum and emission lines with the radio seen in other 
HzRGs at lower redshifts, which can be due to (a combination of) scattered light, emission lines and, possibly, jet-induced 
star formation \citep[e.g.][and references therein]{best98,bicknell00,zirm05}. Several emission lines 
common to high-redshift radio galaxies fall within the \zp\ transmission curve (\civ, \heii). Based 
on a composite radio galaxy spectrum, we estimate that the contribution due to these lines is at most $\sim0.2$ mag in \zp. 
If the continuum is further dominated by the emission of young, hot stars with little dust, we derive a 
star formation rate (SFR) of 13.3 $M_\odot$ yr$^{-1}$. This SFR is comparable to that of normal 
star-forming galaxies at $z\sim4-6$ \citep[e.g.][]{steidel99,papovich01,ouchi04_lf,giavalisco04_results,bouwens04}. 

\subsection{Properties of \lya\ emitting galaxies at $z\approx5.2$}
 
\begin{figure*}[t]
\begin{center}
\includegraphics[scale=.3]{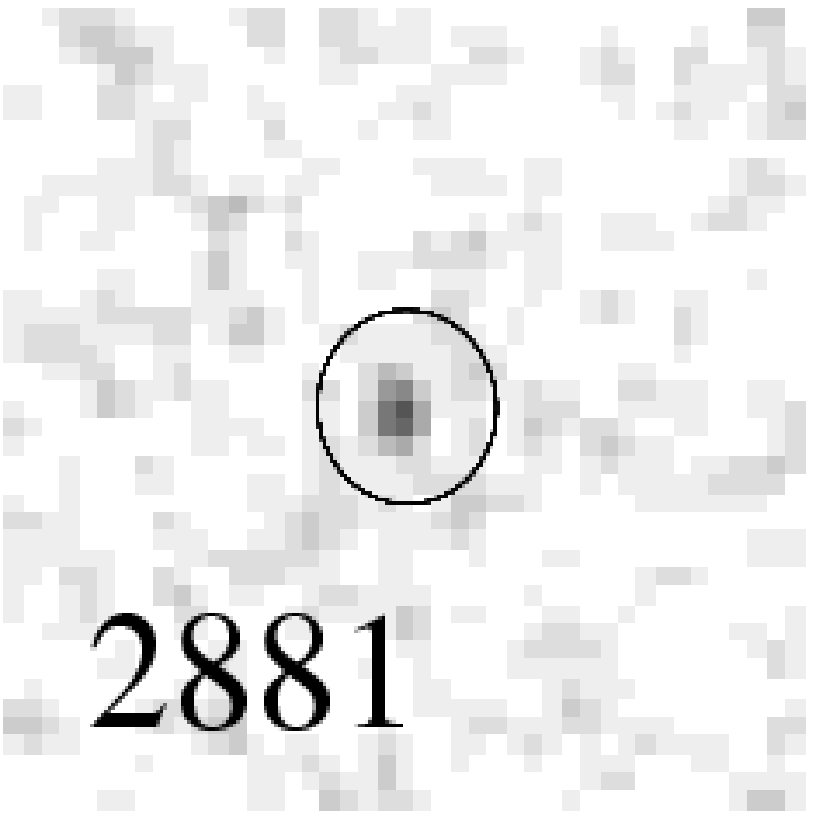}
\includegraphics[scale=.3]{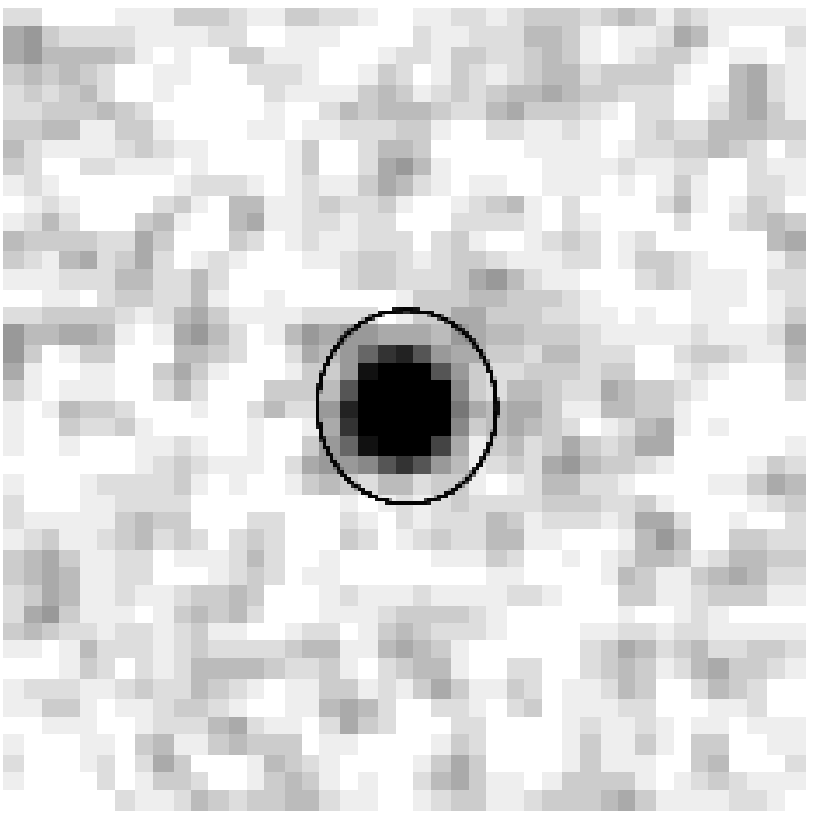}
\includegraphics[scale=.3]{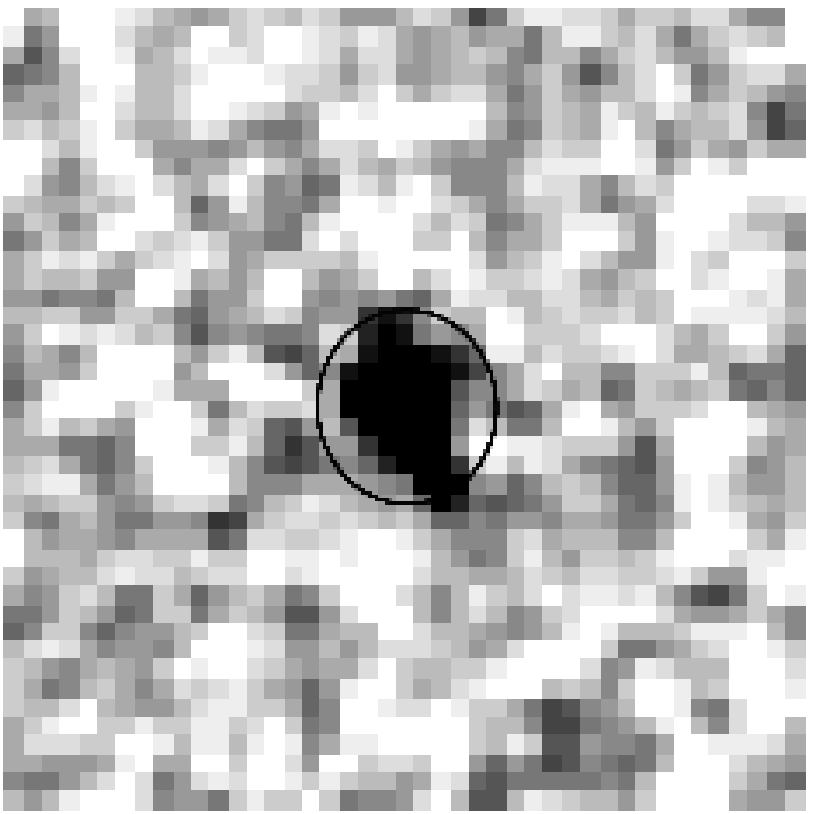}\\
\includegraphics[scale=.3]{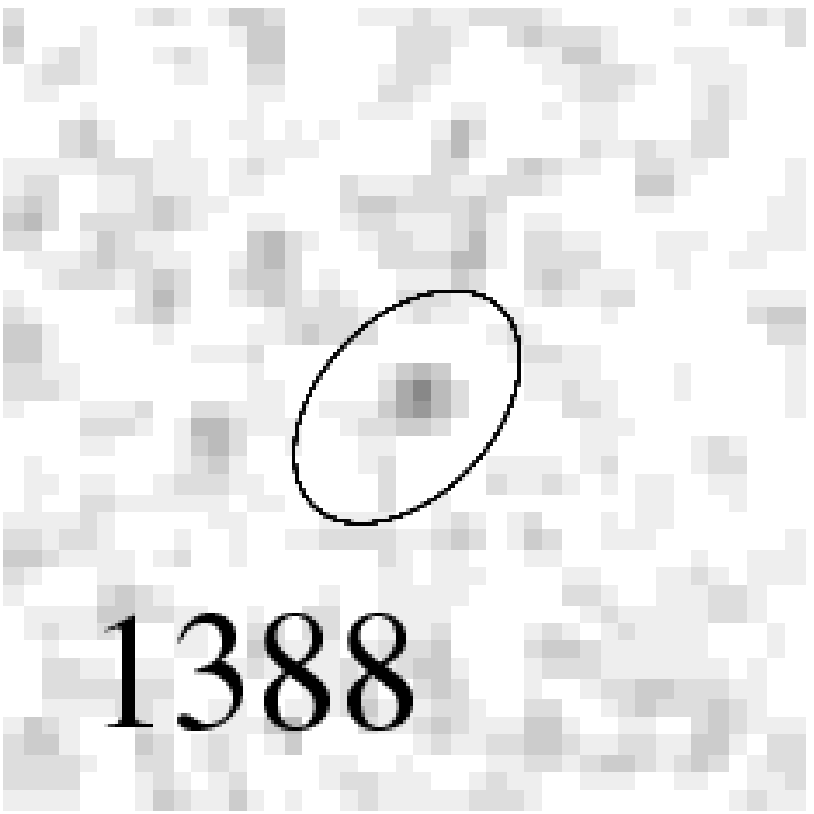}
\includegraphics[scale=.3]{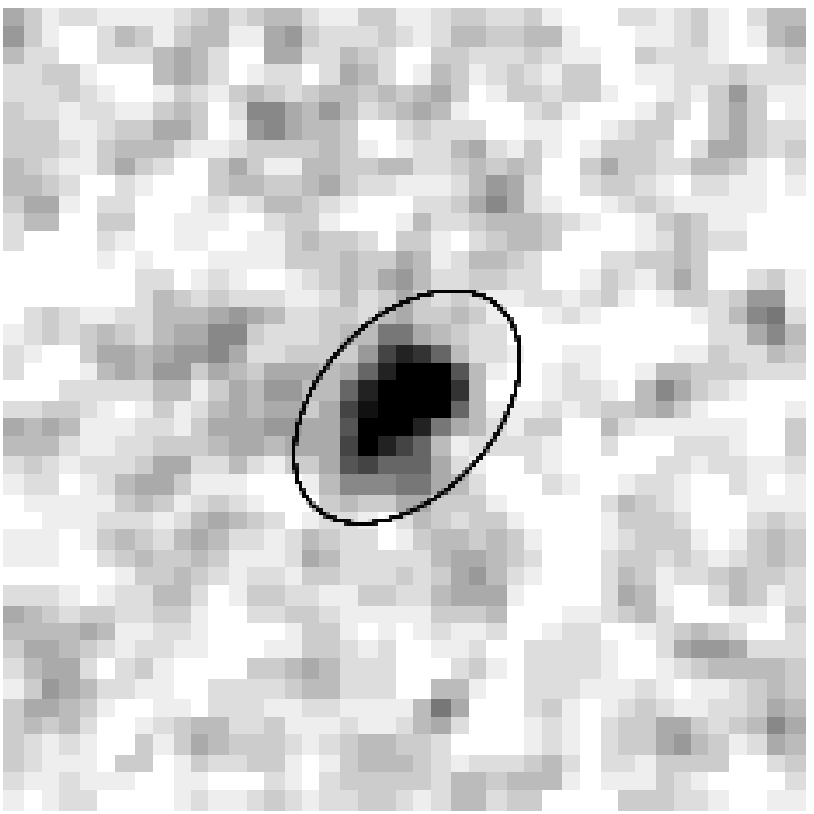}
\includegraphics[scale=.3]{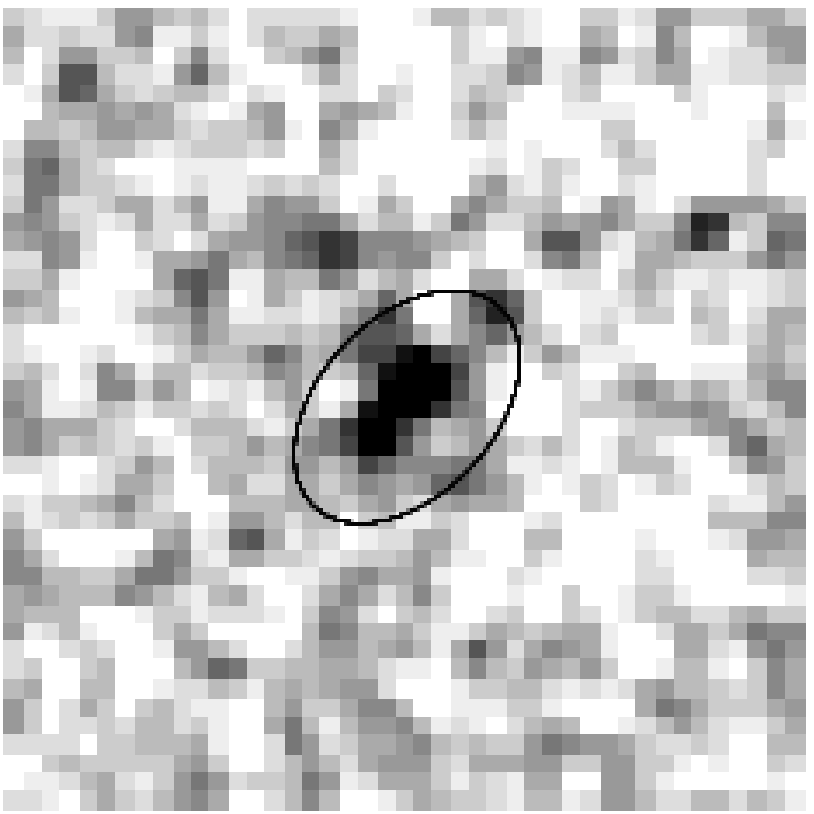}\\
\includegraphics[scale=.3]{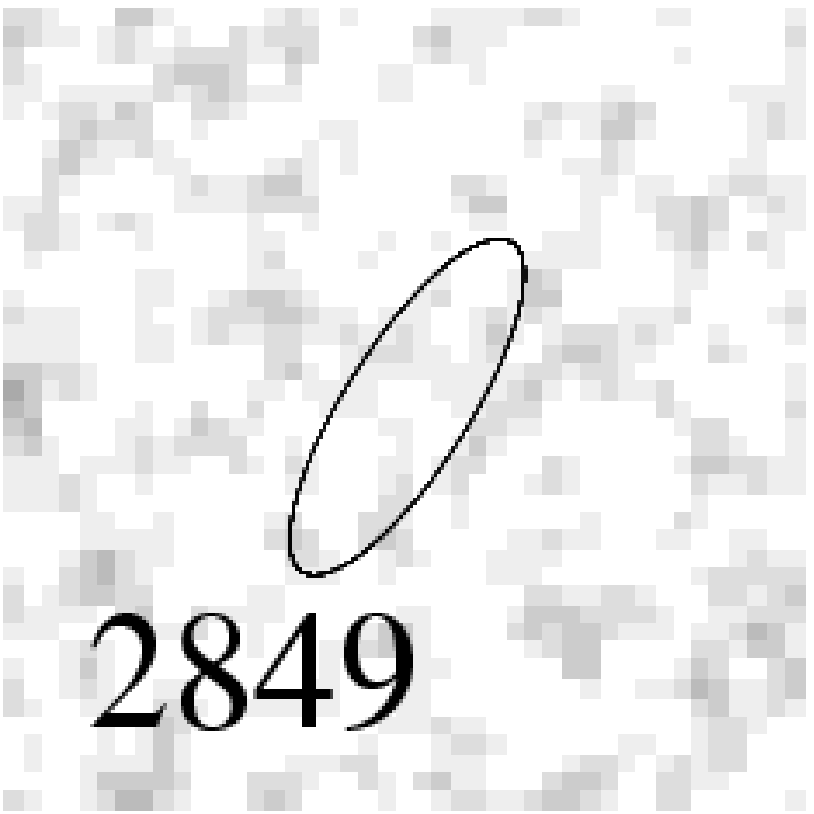}
\includegraphics[scale=.3]{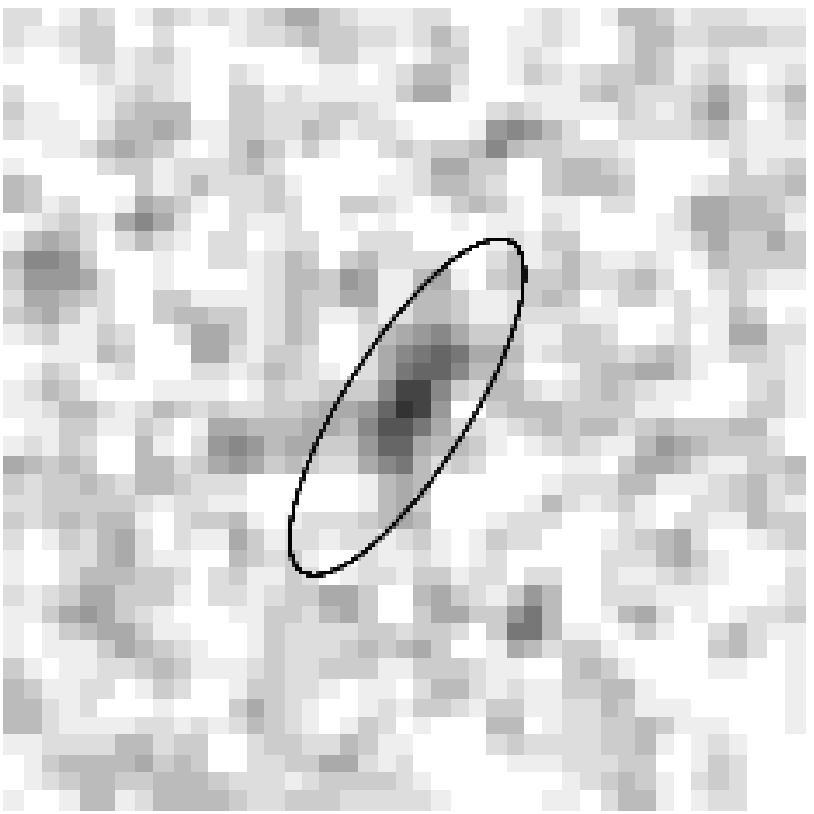}
\includegraphics[scale=.3]{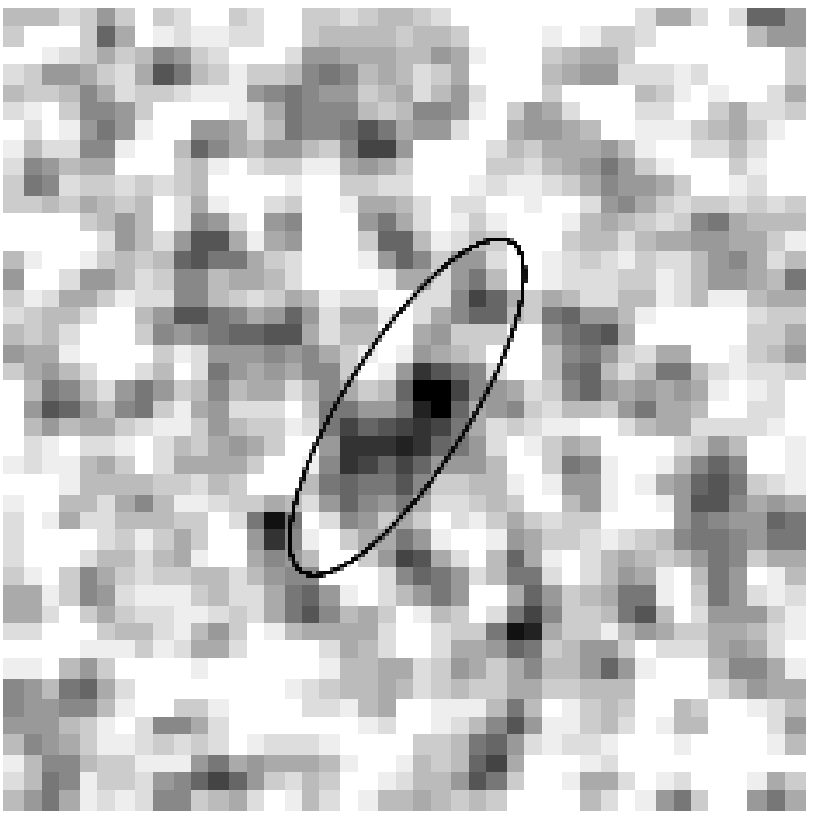}\\
\includegraphics[scale=.3]{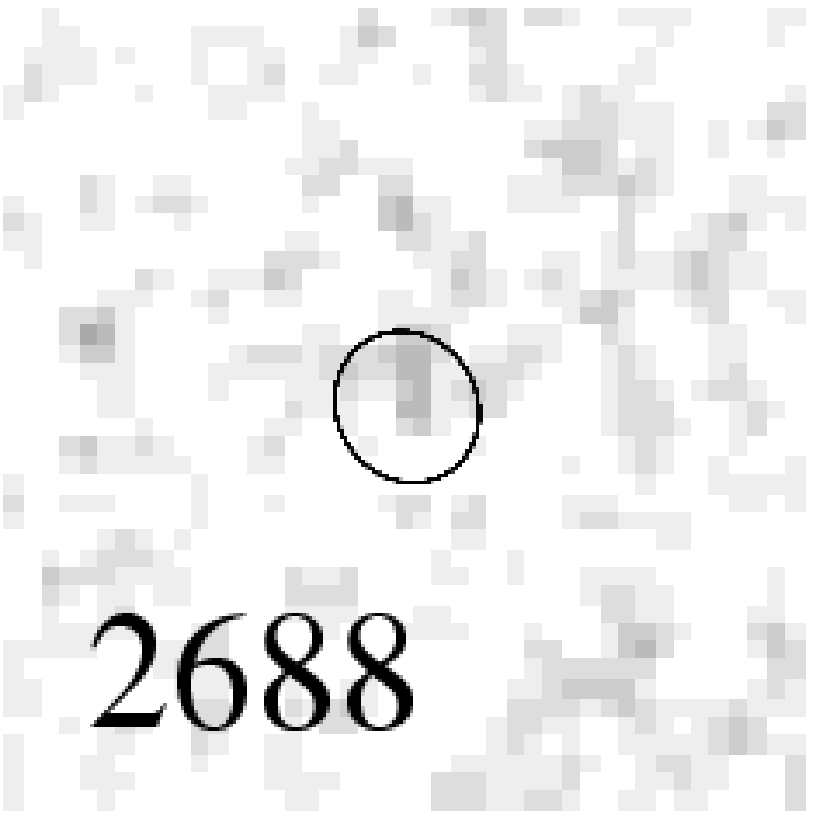}
\includegraphics[scale=.3]{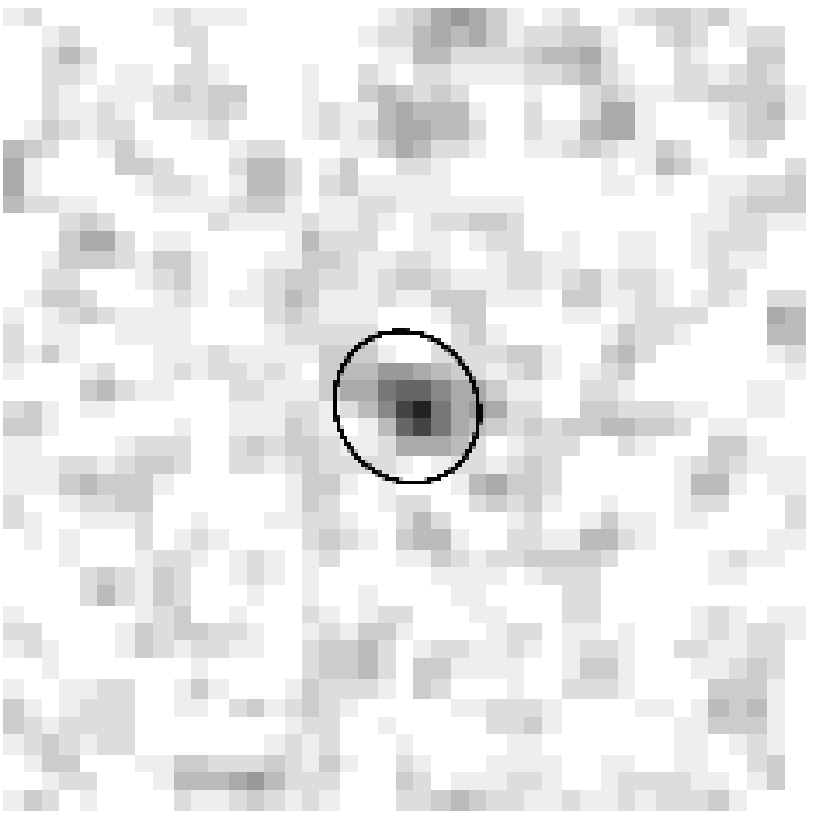}
\includegraphics[scale=.3]{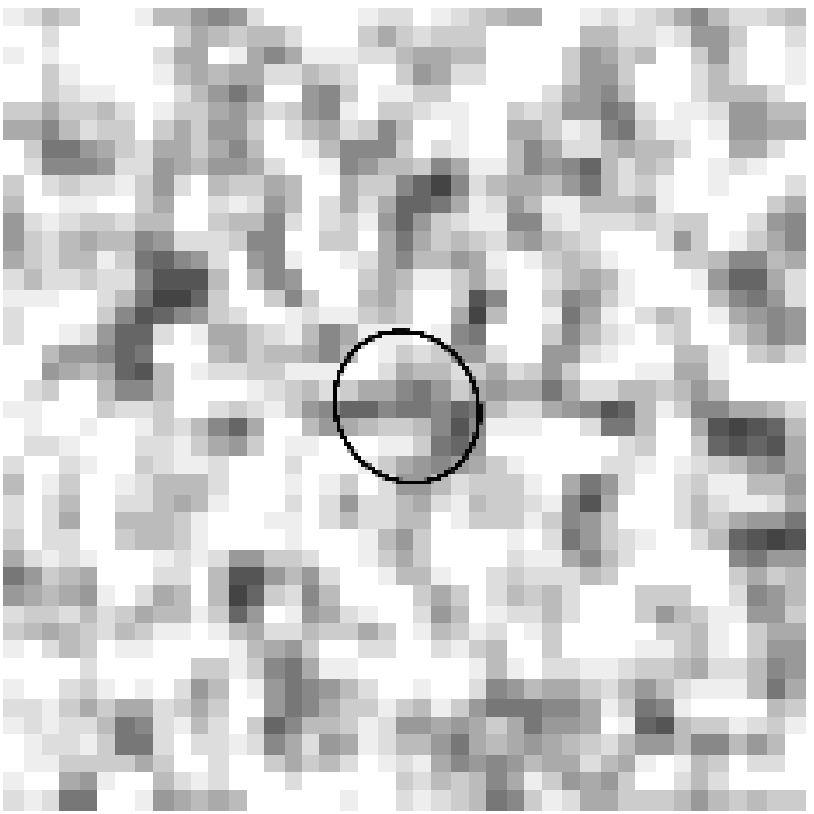}\\
\end{center}
\caption{\label{fig:lya}\vp,\ip\ and \zp\ (from left to right) images of the four spectroscopically confirmed 
\lya\ emitters of \citet{venemans04}. The images have been smoothed using a Gaussian kernel of 0.075\arcsec\ (FWHM). 
Kron apertures determined from the \ip-band image are indicated. The images are $2\arcsec\times2\arcsec$ in size.}
\end{figure*}

In this section we will study some of the properties of the four \lya\ emitting galaxies from \citet{venemans04}.     
The morphologies 
in the 3 bands are shown in Fig. \ref{fig:lya}, and their photometric properties are summarized in Table \ref{tab:laes}. 
All four \lya\ emitters were detected in \ip, the filter that includes \lya, with one object (\#2688) being solely detected in this filter. 
The UV continuum magnitudes measured 
from the \zp-band are all fainter than 25.8 magnitudes, making them fainter than the faintest galaxies in the $z\sim5$ GOODS 
LBG sample from \citet{ferguson04}. This implies that this population of \lya\ galaxies is confined to luminosities of $\lesssim0.7~L^*$, where $L^*$ is the characteristic continuum luminosity of $z=3$ LBGs from \citet{steidel99}. Two emitters have a luminosity of $\lesssim0.3~L^*$.
It is 
evident that the selection of these \lya\ galaxies is biased in two important ways. One, the sample is naturally biased towards 
galaxies with high equivalent width of \lya, and second, it is biased towards the fainter end of the Lyman break galaxy luminosity 
funtion. This finding seems consistent with that of \citet{shapley03} who found evidence that \lya\ equivalent width increases 
towards fainter continuum magnitudes in their spectroscopic $z\sim3$ LBG sample. The faint UV continuum of these \lya\ emitting galaxies is similar to that observed for \lya\ galaxies associated with other radio galaxies \citep[][Overzier et al., in prep.]{venemans04,miley04}.

\subsubsection{Continuum slopes}

 We can use the accurate ACS photometry together with the narrow-band \lya\ flux densities to 
try to sharpen the constraints on the \lya\ EW and to determine the continuum slope ($f_\lambda\propto\lambda^\beta$) of these emitters. 
To this end, we follow the procedures detailed in \citet{venemans05} to subsequently derive the UV slope $\beta$, the strength of the 
continuum, the contribution of \lya\ to \ip, and its (rest-frame) equivalent width, $EW_0$. We take into account that for a source at $z=5.2$ a fraction $Q_{775}\approx0.68$ of the \ip\ flux is absorbed by intervening neutral hydrogen \citep{madau95}, and note that this fraction is 
virtually independent of $\beta$. The uncertainties on $\beta$ and $EW_0$ were obtained by propagating the individual errors on the 
measured magnitudes using a Monte Carlo method, and fitting the resulting distributions with a Gaussian. For object \#2881, the brightest 
in our sample, we find good constraints on both the UV slope and the \lya\ EW, $\beta=-0.8\pm0.6$ and $EW_0=39\pm7$. 
The continuum seems redder than the average slope of $\beta=-1.8\pm0.2$ of \vp-dropouts in GOODS found by \citet{bouwens05_goods}. 
However, we can not rule out the possibility that the \lya\ flux in \ip\ has been over-subtracted due to the 
presence of faint, extended \lya\ in the VLT narrow-band \lya\ image (PSF of $\sim0\farcs8$) not detected with ACS 
(PSF of $\sim0\farcs1$). This could have caused the slope calculated above to be shallower than it in fact is.
We were not able to place tight constraints on the two fainter objects \#1388 and \#2849 detected in \ip\ and \zp, 
and refer to \citet{venemans05} who find $EW_0\sim50$ \AA\ with large errors under the assumption of a flat 
(in $f_{\nu}$, i.e., $\beta=-2$) spectrum.

\subsubsection{Star formation rates}

Using the emission-line free UV flux at 1500\AA\ measured in \zp, we derive star formation rates using the 
conversion between UV luminosity and SFR for a Salpeter initial mass function (IMF) given in \citet{madau98}:
\begin{equation}
\mathrm{SFR} = \frac{L_{1500 \AA} [\mathrm{erg~s}^{-1} \mathrm{Hz}^{-1}]}{8\times10^{27}} M_\odot~\mathrm{yr}^{-1}
\end{equation}
We find 5.9 and 3.0 M$_\odot$ yr$^{-1}$ for objects \#1388 and \#2849. Object \#2881 has a SFR of 9.7 M$_\odot$ yr$^{-1}$, 
quite comparable to that derived for the radio galaxy (see Sect. \ref{sec:rg}). These SFRs are considered to be lower limits, 
since the presence of dust is likely to absorb the (rest-frame) UV luminosities observed. 
The \lya-to-continuum SFR ratios are in the range 0.7--3. The \lya\ SFRs were derived following the standard assumption 
of case B recombination, valid for gas that is optically thick to \hi\ resonance scattering \citep{venemans05}. 

As modeled by \cite{charlot93}, high equivalent width \lya\ is expected for a relatively brief period in young ($\sim10^{7-9}$ yr), 
nearly dust-free galaxies. However, the general understanding is that, regardless of the effects of dust, the UV continuum is a better 
probe of the SFR than \lya, given the large cross-section to resonance scattering for the latter. 
While the \lya\ profile can be severely diminished depending on the geometry of the system, the gas 
density, and the dust contents, the {\it enhancement} of \lya\ flux over UV flux is also not ruled out, at least theoretically. 
Young galaxies may consist of a 2-phase medium \citep[e.g.][]{rees89} effectively thin to \lya\ photons 
scattering off the surfaces of clouds that are optically thick to unscattered UV photons \citep{neufeld91}. 
For the \lya\ emitters found in overdensities associated with radio galaxies we find 
that the star formation rates derived from the UV and \lya\ are generally of a similar order of magnitude 
\citep[e.g., this paper,][Overzier et al., in prep.]{venemans05}. It is unlikely that geometry, dust and scattering medium 
all conspire so that the SFRs derived from \lya\ and the continuum will be comparable. More likely it implies 
that both the UV and \lya\ offer a relatively clear view (e.g. little dust and simple geometry) towards 
the star-forming regions of these galaxies.

\subsubsection{Sizes} 

Except for source \#2688, which we will discuss in detail below, the sources are (slightly) resolved 
in \ip\ and \zp. The half-light radii measured in 
\zp\ are 0\farcs10--0\farcs16, implying that the (projected) physical half-light 
diameters are $<$2.5 kpc, where we have applied a correction for the degree to which half-light radii as 
measured by SExtractor are underestimated for objects with \zp$\approx26$ based on profile simulations 
(Overzier et al., in prep). The sizes are comparable to the sizes we have measured for \lya\ emitters 
associated with radio galaxies at $z=3.13$ and $z=4.11$ \citep[][Overzier et al. in prep.]{miley04,venemans05}. 
We find no evidence for dominant active nuclei among these \lya\ emitters. 

\subsection{A galaxy without UV continuum}

\begin{figure}[t]
\begin{center}
\includegraphics[scale=0.5]{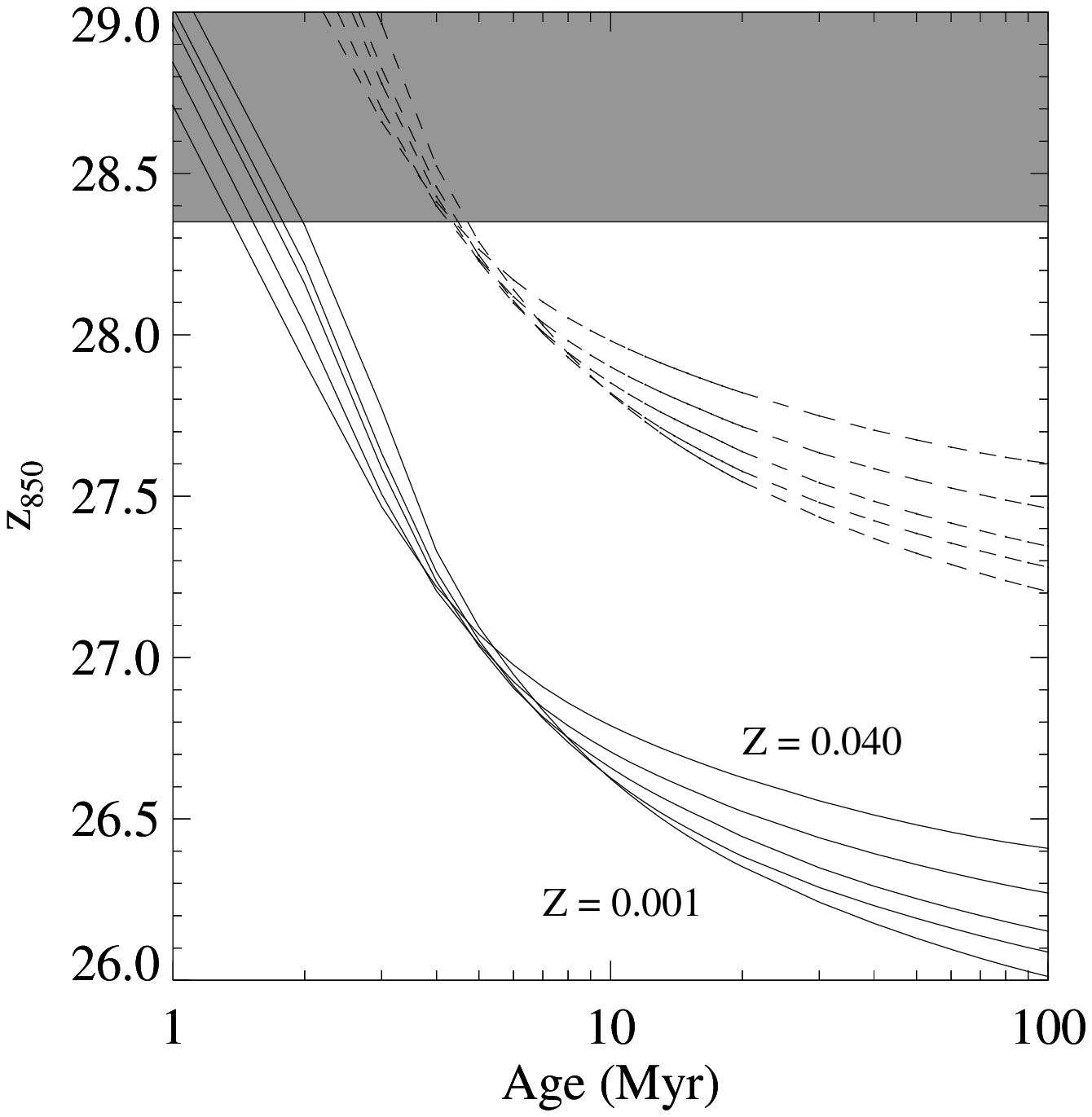}
\end{center}
\caption{\label{fig:sb99}The \zp\ continuum magnitude as a function of age for a 
continuous star formation model with a SFR of 3 $M_\odot$ yr$^{-1}$ from STARBURST99 \citep{leitherer99} (Salpeter IMF with $M_l=1~M_\odot$, $M_u=100~M_\odot$). The metallicities of the models are 0.001, 0.004, 0.008, 0.02, and 0.04 (from bottom to top). 
The shaded box demarcates the $2\sigma$ limit on the \zp\ magnitude within the Kron aperture of \lya\ emitter \#2688. 
Dashed lines indicate similar models, but with a SFR of only 1 $M_\odot$ yr$^{-1}$. The non-detection in \zp\ may indicate that \#2688 
has an age of only a few Myr.}  
\end{figure}
 
Object \#2688 from \citet{venemans04} is particularly interesting. It is the 
faintest object in our \lya\ sample (\ip$\approx28$) and it is not detected 
in \zp\ at the $2\sigma$ level (\zp$>28.4$). Likewise, there is no detection in \vp. 
Assuming $\beta\approx-2$, which is appropriate for a dustless, 
young (1--100 Myr) galaxy, correcting for the \lya\ emission in the \ip-band
would place this object's magnitude close to the 
detection limit in \ip. This implies that the \ip\ flux is solely that of \lya, with an  
$EW_0$ of $>100$ \AA. What physical processes could explain its peculiar observed 
properties?

$\bullet$ {\bf Young star-forming galaxy?} If object \#2688 is a young star-forming galaxy, the observed equivalent width of 
\lya\ should be a function of the age of the stellar population.  
\citet{venemans04} estimated the star formation rate in \#2688 from its \lya\ luminosity and found $\sim3~M_\odot$ yr$^{-1}$.  
Based on a population synthesis model for a 
young stellar population with a SFR of 3 $M_\odot$ yr$^{-1}$ (Salpeter IMF, $M_l=1M_\odot$, $M_u=100M_\odot$) shown in 
Fig. \ref{fig:sb99}, our robust limit on the \zp-band ($\sim1465$ \AA\ rest-frame) would be surpassed 
within only $\sim 2$ Myr \citep{leitherer99}.  
A comparably young object was found by \citet{ellis01} at $z=5.6$. In this case, the lensing 
amplification by the cluster Abell 2218 also enabled to place a strong upper-limit on the 
object size of 150 pc, consistent with it being a typical \hii\ region.  
Given the non-detection of \#2688 in \zp, it is difficult to place an upper limit on the size of the object. If we take 
the physical half-light diameter of 2.5 kpc derived for the other \lya\ emitters as an extreme upper limit on the 
size, it is not likely that star formation can progress over such large size within only a few Myr. Whether the actual 
size of the \lya\ emitting in \#2688 is similar to that of typical \hii\ regions is unclear given the extremely thin detection in \ip.

$\bullet$ {\bf Outflow?} Extended emission line regions seem to be a common feature of both local and high-redshift star-forming galaxies. 
Locally, some of the emission is produced in galactic scale outflows.  
Empirically there is a lower limit to the surface density of star formation necessary to 
launch such a galactic wind of $0.1~M_{\odot}$ yr$^{-1}$ kpc$^{-2}$ \citep{heckman90}.  
For \#2688 we calculate a star formation rate surface density of $>0.5~M_\odot$ yr$^{-1}$ kpc$^{-2}$.  Therefore, 
the observed high equivalent-width \lya\ may be a result of outflowing gas, while the galaxy itself may 
be obscured and older than the strong upper limit of a few Myr derived for the above scenario.

$\bullet$ {\bf AGN?} The equivalent width of \lya\ could also be boosted by the presence of an active 
nucleus.  TN J0924--2201 itself has a \lya\ $EW_0 = 83$ \AA\, close to the lower limit derived 
for \#2688.  While there is no evidence for a bright nuclear point-source in \#2688, it could 
be easily obscured by circumnuclear dust, particularly at rest-frame ultraviolet 
wavelengths. Because the spectrum only has narrow \lya, it could be a faint narrow line quasar. 
Several of the \lya\ emitters in the protocluster near radio galaxy MRC 1138--262 at $z=2.16$ 
have been detected with {\it Chandra} indicating that the AGN fraction of such protoclusters 
could be significant. In contrast, \citet{wang04} found no evidence for AGN among a large 
field sample of $z\approx4.5$ \lya\ emitters observed in the X-ray. 

\subsection{Selection of \vp-dropouts}
\label{sec:selection}

Galaxies without a significant excess of \lya\ (i.e. rest-frame $EW_{Ly\alpha}<20$\AA) constitute $\sim75$\% of LBG samples \citep[][]{shapley03}, and hence are missed by selection purely based on the presence of \lya\ 
emission. To circumvent this inherent bias in \lya\ surveys, galaxies can be selected on the basis of broad-band colors that 
straddle the Lyman break for some specific redshift range. Unfortunately, having only a few filters, the Lyman break selection 
provides only a crude selection in redshift space due to photometric scatter and uncertainty in the underlying spectral energy 
distributions. This is especially important when we want to test for the presence of LBGs within a relatively narrow 
redshift range of the radio galaxy. 

\citet{giavalisco04_results} selected \vp-dropouts from the GOODS fields using the criteria:
\begin{eqnarray}
\label{eq:criteria1}
&&[(V_{606}-i_{775}) \ge 1.5+0.9\times(i_{775}-z_{850}) ~\vee\nonumber\\
&&(V_{606}-i_{775}) \ge 2.0]~\wedge~ (i_{775}-z_{850})\le1.3 ~\wedge\nonumber\\
&& (V_{606}-i_{775}) \ge 1.2
\end{eqnarray}
where $\vee$ and $\wedge$ are the logical OR and AND operators.  
Although we will use these selection criteria to select \vp-dropout samples from our datasets, we will use 
a slightly modified selection window when discussing the clustering statistics of \vp-dropouts with respect 
to the radio galaxy (\se\ref{sec:overdensity}). 
We can tighten the color constraints given in Eq. \ref{eq:criteria1} to effectively remove 
relatively blue objects that are likely to be at redshifts much lower than we are interested in ($z\approx5.2$), 
as well as relatively red objects at much higher redshifts. We required
\begin{equation}
\label{eq:criteria2}
0.0\le(i_{775}-z_{850})\le1.0
\end{equation}
in addition to Eq. \ref{eq:criteria1} to reject galaxies at $z\lesssim4.8$ and $z\gtrsim5.5$, based on the 
color-color track of a $10^8$ yr constant star forming model of $0.4Z_\odot$ metallicity. The resulting selection window 
is indicated in Fig. \ref{fig:cc} (shaded area). The selection window of \citep{giavalisco04_results} as given in Eq. \ref{eq:criteria1} 
has been indicated for comparison (dashed line). 

Unlike GOODS and the UDF parallel fields, there are no 
observations in \bp\ for our field, which makes it impossible to remove low redshift contamination 
by requiring a maximum upper limit on detections in \bp\ (e.g., $S/N<2$). The estimates for the low 
redshift contamination fraction of \vp-dropouts from GOODS amount to $\sim10-30$\% \citep[][]{bouwens05_goods}. 
We note, however, that in some cases low redshift objects that have made it into the selection window 
can still be rejected on the basis of their high relative brightness and/or large sizes in the \vp\ip\zp-bands 
during visual inspection.  

\subsection{Properties of \vp-dropouts in the field of TN J0924--2201}

\begin{figure}[t]
\begin{center}
\includegraphics[width=0.5\textwidth]{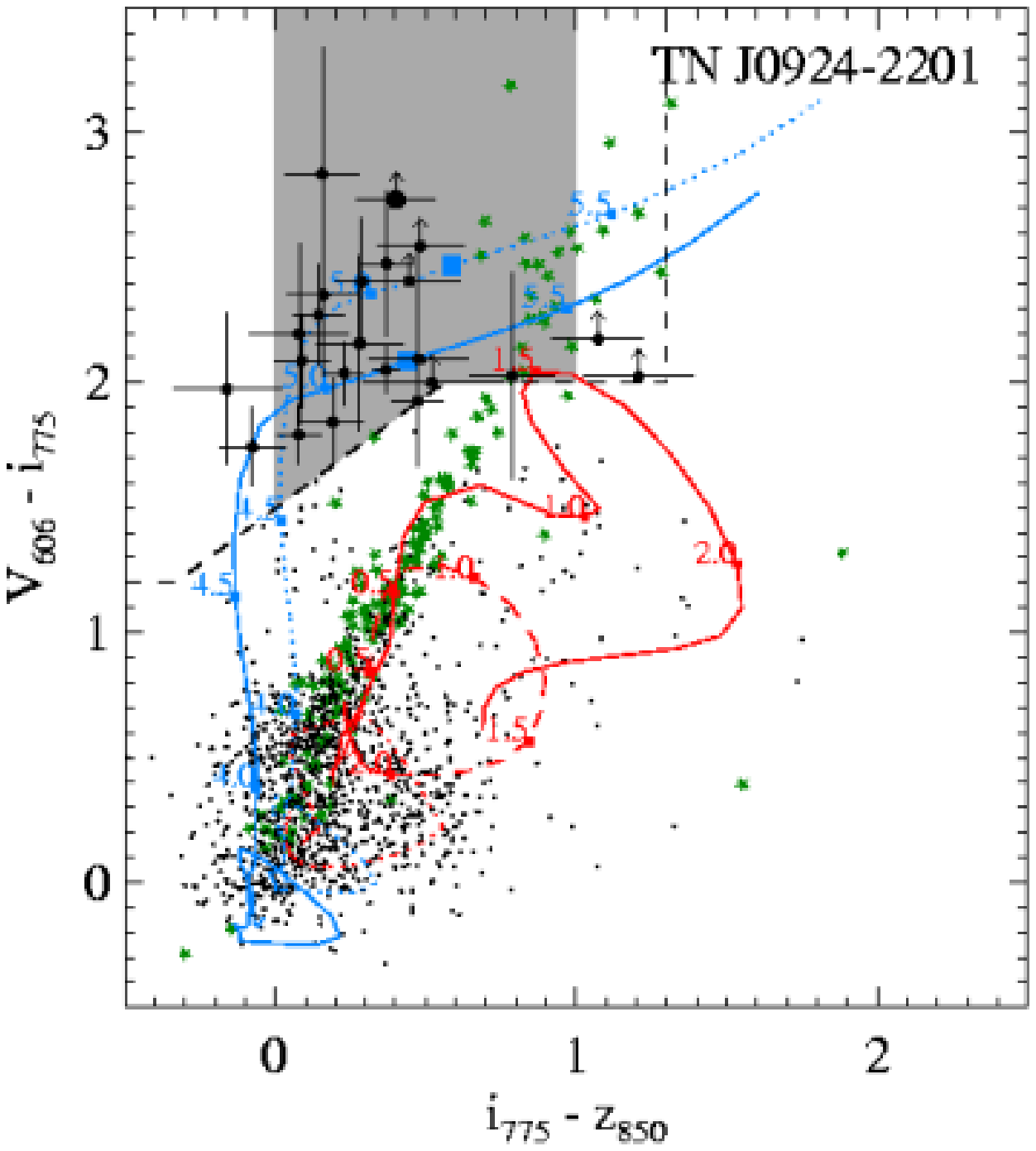}
\end{center}
\caption{\label{fig:cc}\vp--\ip\ vs. \ip--\zp\ color-color diagram for TN J0924--2201 candidate LBGs  
(solid circles) relative to objects from the full catalogue (\zp$<26.5$ with $S/N>5$). 
The radio galaxy is indicated by the large circle. The black dashed line is the selection window of \citet{giavalisco04_results}. 
The shaded region marks the selection window used for 
the clustering statistics of \vp-dropouts at $z\approx5.2$ (see \se\ref{sec:overdensity}). 
The spectral tracks are from an elliptical (red solid line), a Sbc (red dashed line), a Scd (red dotted line), 
and a 100 Myr constant star formation model with $E(B-V)=0.0$ (blue solid line) and $E(B-V)=0.15$ (blue dotted line).
Redshifts are indicated along the tracks. The redshift of the overdensity of \citet{venemans04} 
 is marked by blue squares ($z=5.2$). 
The position of the stellar locus is illustrated by the green stars (objects having $S/G>0.85$). 
All colors were set to their $2\sigma$ limits (limits and error bars for field objects have been omitted for clarity).}
\end{figure}

\begin{figure}[t]
\begin{center}
\includegraphics[width=0.5\textwidth]{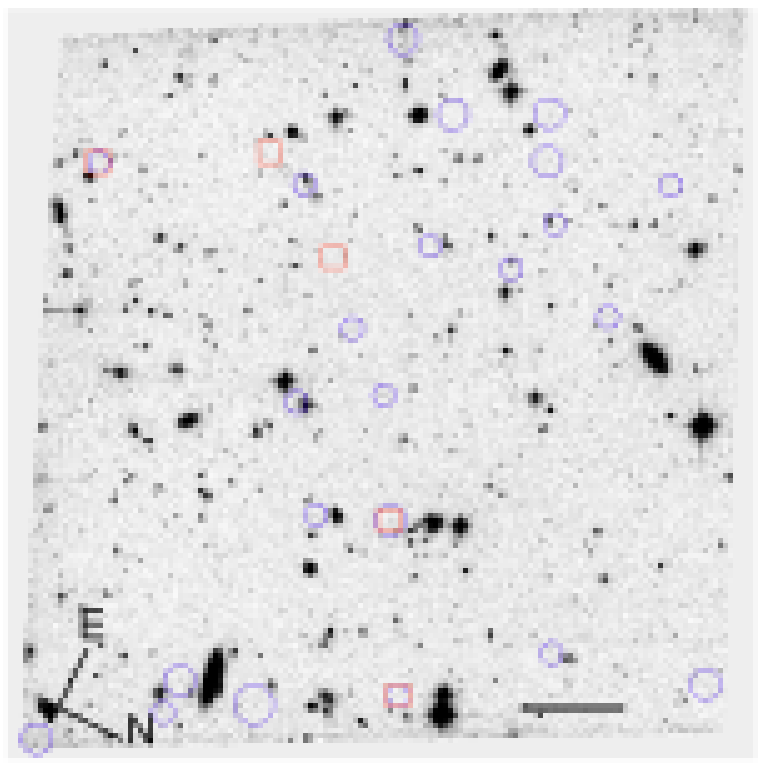}
\end{center}
\caption{\label{fig:image}ACS \zp\ image showing the positions of the \vp-break objects with blue circles. The size of the 
circles scales with the total \zp\ magnitude of the LBGs, where the smallest circles correspond to objects with \zp$>25.5$, 
and the largest correspond to objects brighter than \zp$<24.5$. The positions of the spectroscopically confirmed \lya\ emitters 
from \citet{venemans04} are indicated with red squares. TN J0924--2201 is located roughly 0.5\arcmin\ from the image center 
towards the bottom of the image (this is both a \lya\ emitter and a \vp-break object). The scale bar at the bottom measures 0\farcm5.}
\end{figure}

The \vp--\ip\ versus \ip--\zp\ diagram of the objects that meet our selection criteria is shown in 
Fig. \ref{fig:cc} compared to the entire \vp\ip\zp\ sample. Also shown are the color-color tracks 
of several standard SEDs and the stellar locus. We find 23 \vp-dropouts 
down to a limiting magnitude of \zp$=26.5$.  The radio galaxy (\#1396) and the two brightest \lya\ emitters 
(objects\footnote{\#2881 and \#1388 in Table \ref{tab:laes} and \citet{venemans04}} \#449 and \#1844) 
passed the \vp-dropout selection criteria. Table \ref{tab:lbgs} lists the coordinates, colors and magnitudes of 
the LBG candidates. 
Fig. \ref{fig:image} shows the \zp-band image with the positions of the \vp-dropouts (blue circles) and the \lya\ 
emitters (red squares).

\subsubsection{SFRs and continuum slopes}
 
Our limiting magnitude in \zp\ corresponds to $\sim0.5~L^*$ (taking into account the average 
amount of flux missed).
We calculated star formation rates from the emission-line 
free UV flux measured in \zp. The SFRs range from 5--42 $M_\odot$ yr$^{-1}$ if there is no dust (see Table \ref{tab:lbgs}). 
We calculated an average UV slope ($f_\lambda\propto\lambda^\beta$) for the entire sample from the \ip--\zp\ 
color and find $\langle\beta\rangle=-2.4$ with a standard deviation of 1.7 for the sample.  
Here we have assumed a redshift of $z=5.2$ to 
convert between magnitudes and the actual flux densities of the continuum in \ip. 
However, this assumed redshift is critical to the calculation of $\beta$, due to its large dependence on the amount 
of \lya\ forest absorption in \ip:  the average \ip--\zp\ color corresponds to slopes ranging 
from $\langle\beta\rangle=-1.3$ at $z=5.0$ to $\langle\beta\rangle=-4.0$ at $z=5.4$. The average 
slope of $\langle\beta\rangle=-2.4$ that we measured is consistent with the average slope of \vp-dropouts ($\beta=-1.8\pm0.2$) in 
GOODS \citep{bouwens05_goods}. 

It is impossible to fit both the redshift and the spectral slope independently. In the following we will assume that $z=5.2$, 
and that the value of the spectral slope is largely determined by dust, rather than age or metallicity. We have parametrized 
$E(B-V)-\beta_{iz}$ for a base template consisting of a 100 Myr old ($z_f\approx5.6$) SED with $0.2~Z_\odot$ metallicity that 
has been forming stars at a continuous rate \citep[from][]{bruzualcharlot03}. The template was reddened by applying increasing 
values of $E(B-V)$ using the recipe of \citet{calzetti00}. The measured slopes are consistent with modest absorption by dust of 
$E(B-V)\sim0-0.4$, with the lower values preferred given the mean slope of the sample. In some cases we also found negative values 
of $E(B-V)$. This suggests that the color might be bluer than that of the base template used, or that the redshift is off. 

\citet{bouwens05_goods} found evidence for evolution in the mean UV slope from $z\sim5$ ($\beta=-1.8\pm0.2$) to $z\sim2.5$ ($\beta=-1.4\pm0.1$) \citep[see also][]{lehnert03,ouchi04_lf,papovich04,bouwens05_z6}. They have interpreted this as an evolution in the dust 
content rather than age or metallicity, based on the plausible assumption that any change in these parameters by significantly 
large factors seems unlikely given that the universe only doubles in age over this redshift interval and the gradual process 
of galaxy formation. Reducing the dust content by a factor of $\sim2$ from $z\sim3$ to $z\sim5$ can explain the relatively blue continuum 
of the \vp-dropouts.


\begin{figure*}[t]
\begin{center}
\includegraphics[width=0.7\textwidth]{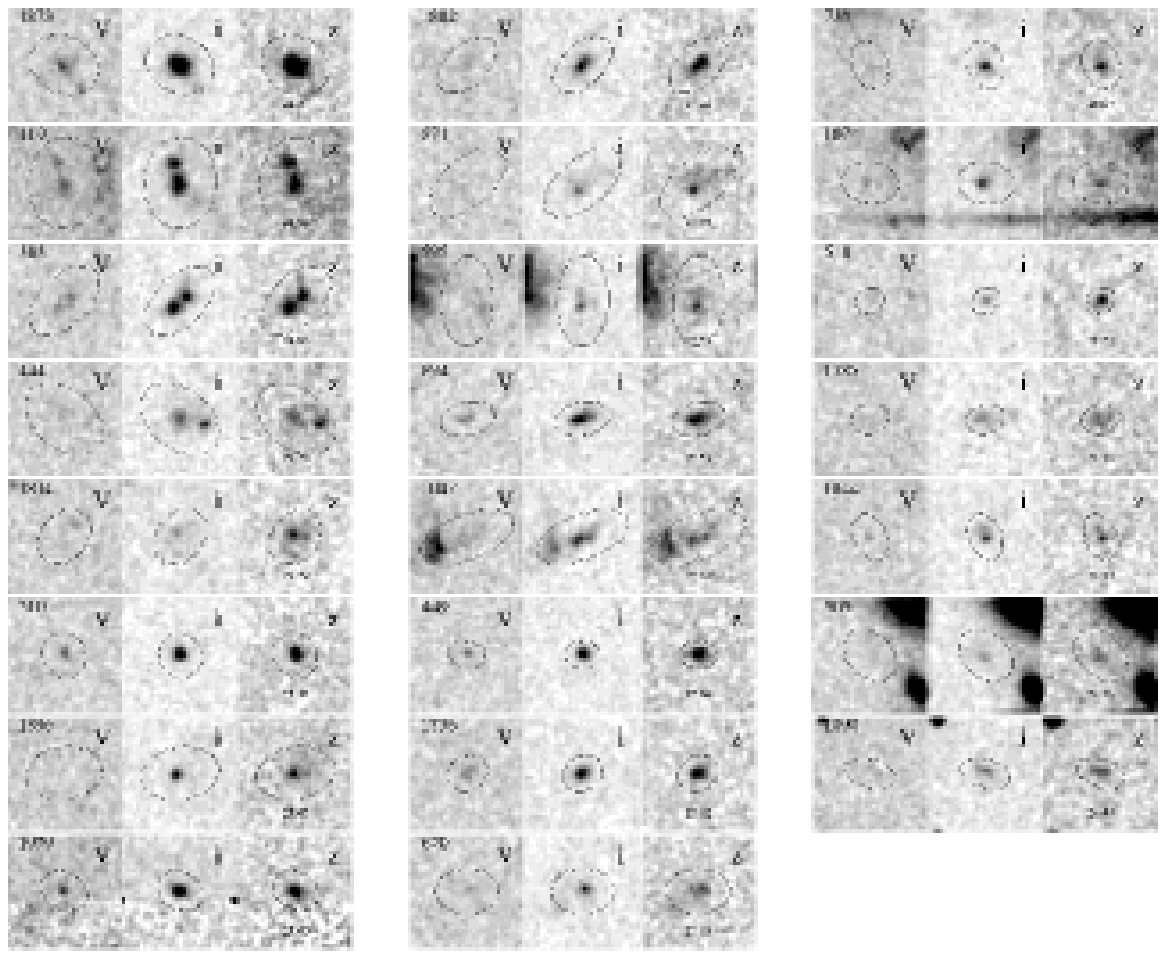}
\end{center}
\caption{\label{fig:stamps}HST/ACS postage stamps showing (from left to right) \vp, \ip, and \zp\ of the $z\sim5$ LBG sample. 
Each image measures $2\arcsec\times2\arcsec$, corresponding to $\sim13\times13$ kpc at $z\sim5$. The Kron aperture defined by the light distribution in \zp\ has been indicated. The images have been smoothed using a Gaussian kernel of 0.075\arcsec\ (FWHM).}
\end{figure*}

\subsubsection{Sizes}

We have measured half-light radii in \zp. A Gaussian fit to the size distribution 
gives a $\langle r_{hl,z}\rangle=0\farcs16$ with standard deviation $0\farcs05$. The mean half-light radius corresponds 
to $\sim1.2$ kpc at $z\sim5$. Note that our sample is biased against $z\sim5$ AGN point sources, since they would be rejected 
based on their high stellarities. If we divide our sample in two magnitude ranges \zp$=24.2-25.5$ and \zp$=25.5-26.5$, 
the mean half-light radii for the two bins are 0\farcs20 and 0\farcs14, respectively. 

While it cannot entirely be ruled out that fainter LBGs are intrinsically smaller, the observed difference between the two bins can most 
likely be explained by the effect of surface brightness dimming in two ways: 1) the fraction of 
light that is missed in aperture photometry is larger for fainter sources, and 2) the incompleteness is higher for 
larger sources at a fixed magnitude \citep[see e.g.][]{bouwens04,giavalisco04_survey}. The mean 
half-light radius of \zp$<25.8$ LBGs at $z\sim5$ in GOODS is $\langle r_{hl,z}\rangle\approx0\farcs27$, as measured 
by \citet{ferguson04}. However, \citet{ferguson04} measured half-light radii using maximum apertures approximately $4\times$ 
larger than ours, which inevitably results in slightly larger half-light radii. Calculating the half-light radius using our 
method and our own sample of $z\sim5$ LBGs from the GOODS field (\se \ref{sec:overdensity}) gives $\langle r_{hl,z}\rangle=0\farcs17\pm0.06$ 
(with $0\farcs20\pm0.08$ and $0\farcs16\pm0.05$ for the brighter and fainter magnitude bins, respectively), consistent 
with the sizes we find in the TN J0924--2201 field.

\begin{figure}[t]
\begin{center}
\includegraphics[width=0.5\textwidth]{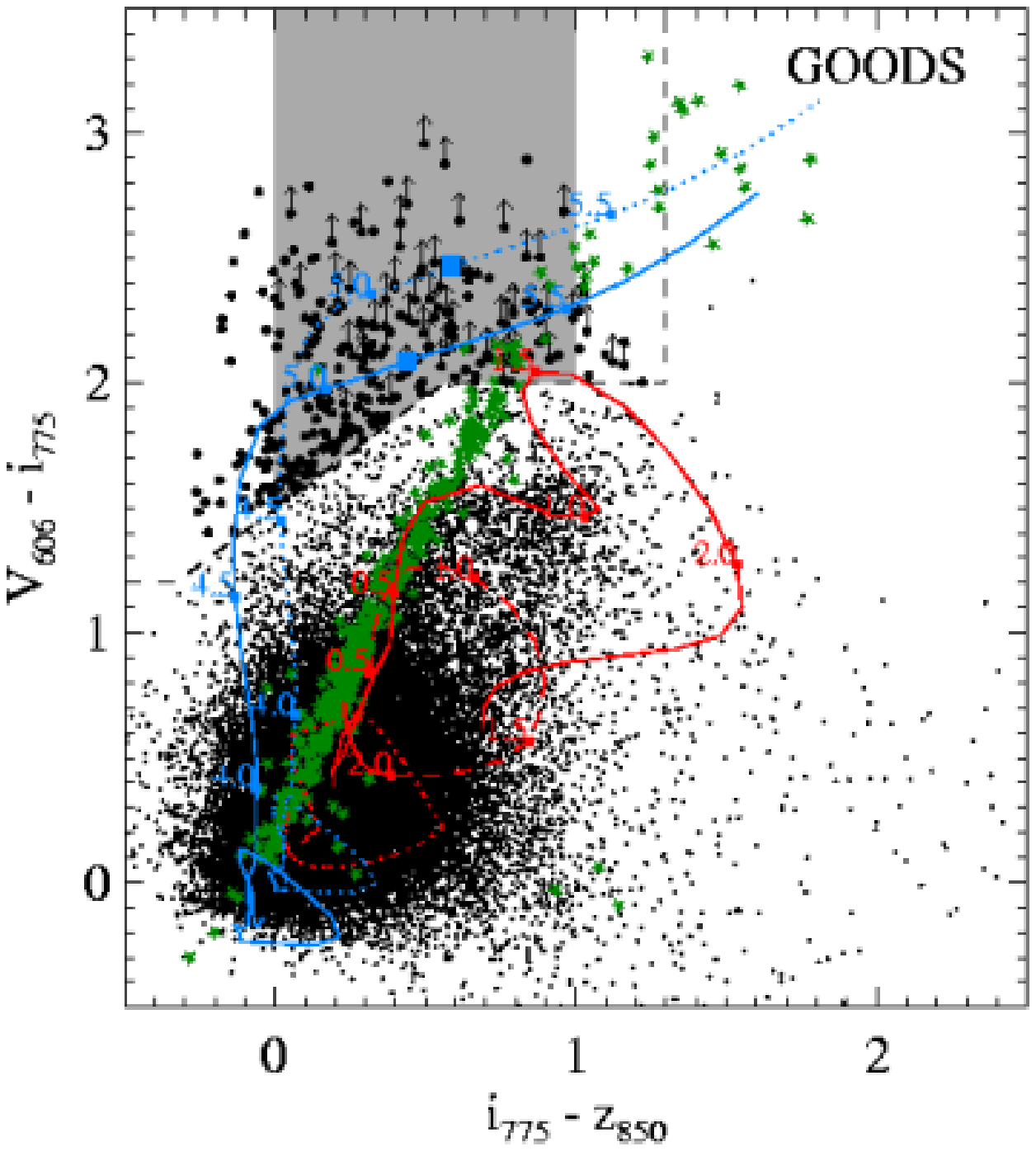}
\end{center}
\caption{\label{fig:cccontrol}Color-color diagram of GOODS. See the caption of Fig. \ref{fig:cc} for details.}
\end{figure}

The \vp, \ip, and \zp\ morphologies are shown in Fig. \ref{fig:stamps}. Three objects (\#119, \#303, and \#444) 
have a clear double morphology. Based on the large \vp-dropout sample from GOODS 
(\se \ref{sec:overdensity}) we would expect roughly 1.5 of such systems in our field, indicating that our field 
might be relatively rich in merging systems. A more detailed, comparative analysis of sizes and morphologies of 
LBGs and \lya\ emitters in radio galaxy protoclusters at $2<z<5.2$ will be given elsewhere.  

\begin{figure*}[p]
\begin{center}
\includegraphics[width=0.5\textwidth]{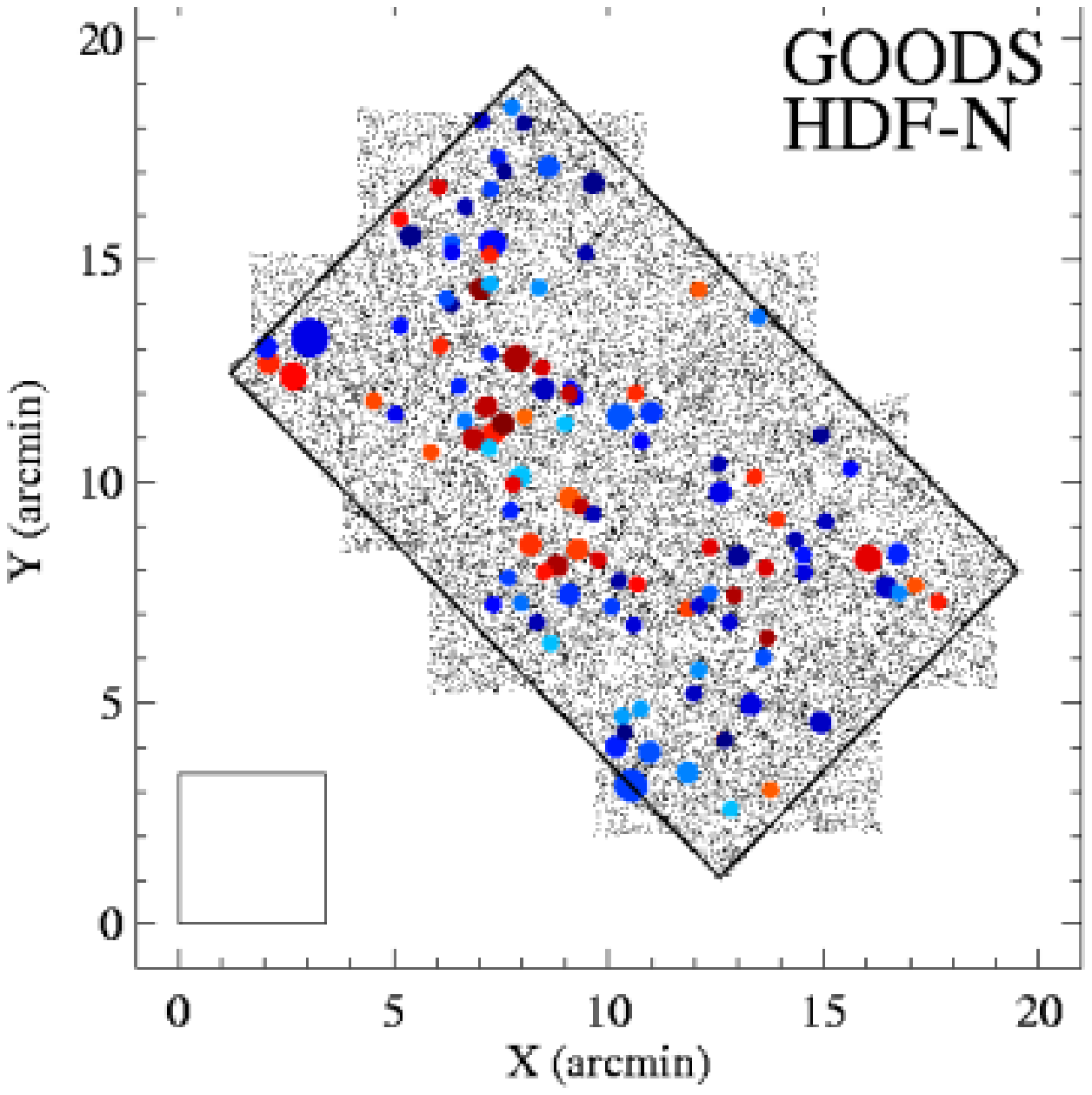}~~~\includegraphics[width=0.5\textwidth]{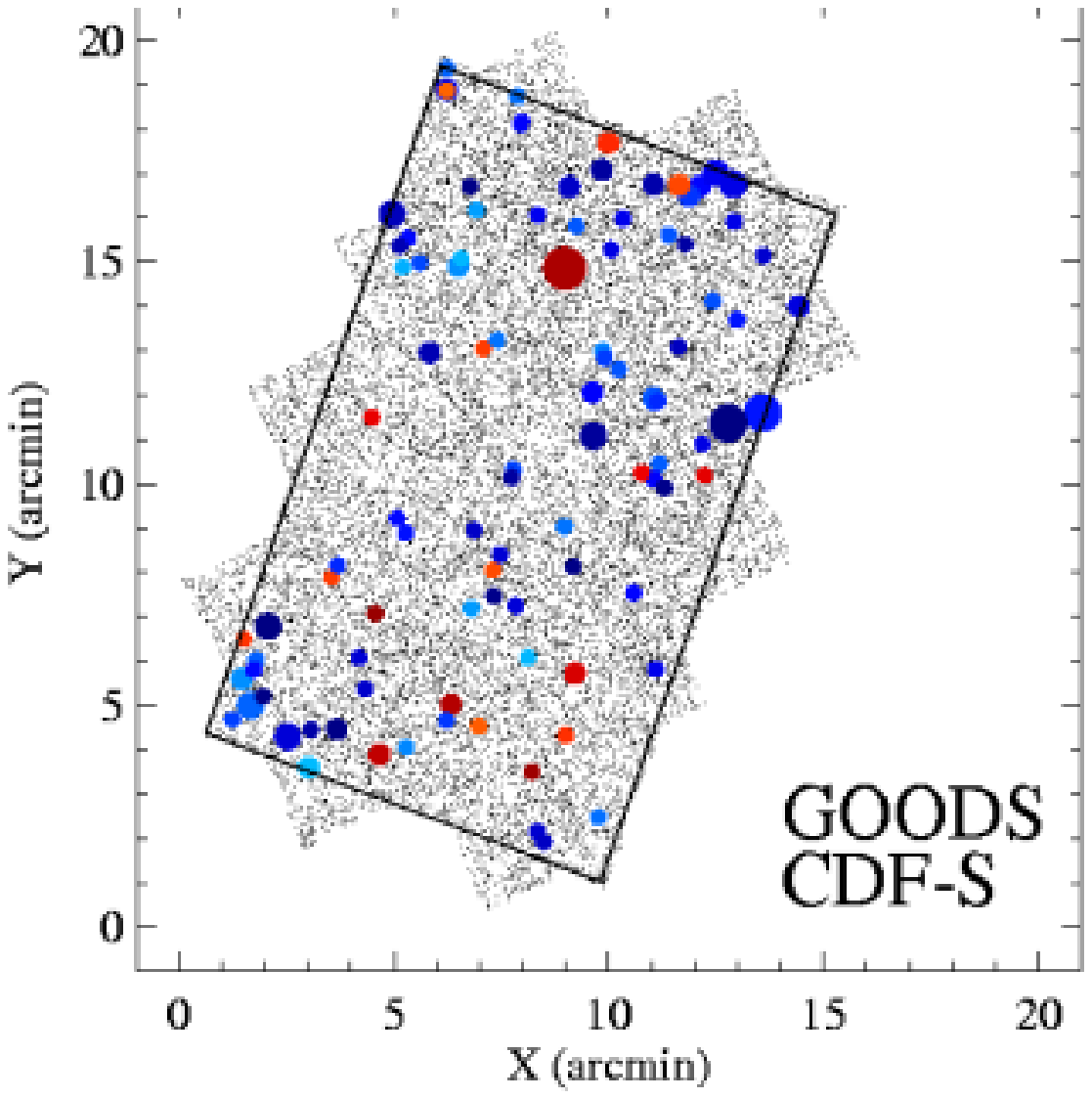}
\end{center}
\caption{\label{fig:2dgoods}Angular distribution of \vp-break objects (solid circles) in the GOODS HDF-N field (left panel) 
and the CDF-S field (right panel), compared to the photometric sample as a whole (points). The \vp-dropouts 
have been color-coded corresponding to their \ip--\zp\ colors (dark blue corresponds to \ip--\zp$\approx0.0$, dark red corresponds to 
\ip--\zp$\approx1.0$), which can be taken as a rough measure for the relative redshifts.
The size of the symbols scales with \zp\ magnitude (the smallest symbols correspond to \zp$>$26.0, the largest symbols to \zp$<$24.0). 
The total area of the GOODS fields is $\sim314$ arcmin$^{2}$ (area within the solid lines). The size of a single $3\farcm4\times3\farcm4$ ACS pointing as obtained for TN J0924--2201 has been indicated to the left of the GOODS HDF-N field for comparison.}
\end{figure*}

\subsubsection{Point sources}

The Galactic stellar locus runs through our \vp-dropout selection window (green stars in Fig. \ref{fig:cc}). 
We found $\sim14$ stellar objects that pass our selection criteria, if we let go of the requirement of relatively low  
stellarity index, as measured by SExtractor. However, the additional objects we found were all brighter than \zp=25.0, 
and the majority were scattered around the red end of the stellar locus. No new objects with high stellarity were found 
at fainter magnitudes. Therefore, we believe that we have not missed a significant population of (unresolved) $z\sim5$  
AGN in this field.  

\section{Discussion}

\subsection{An overdensity of \vp-dropouts associated with TN J0924--2201?}
\label{sec:overdensity}

\begin{figure}[t]
\begin{center}
\includegraphics[width=0.5\textwidth]{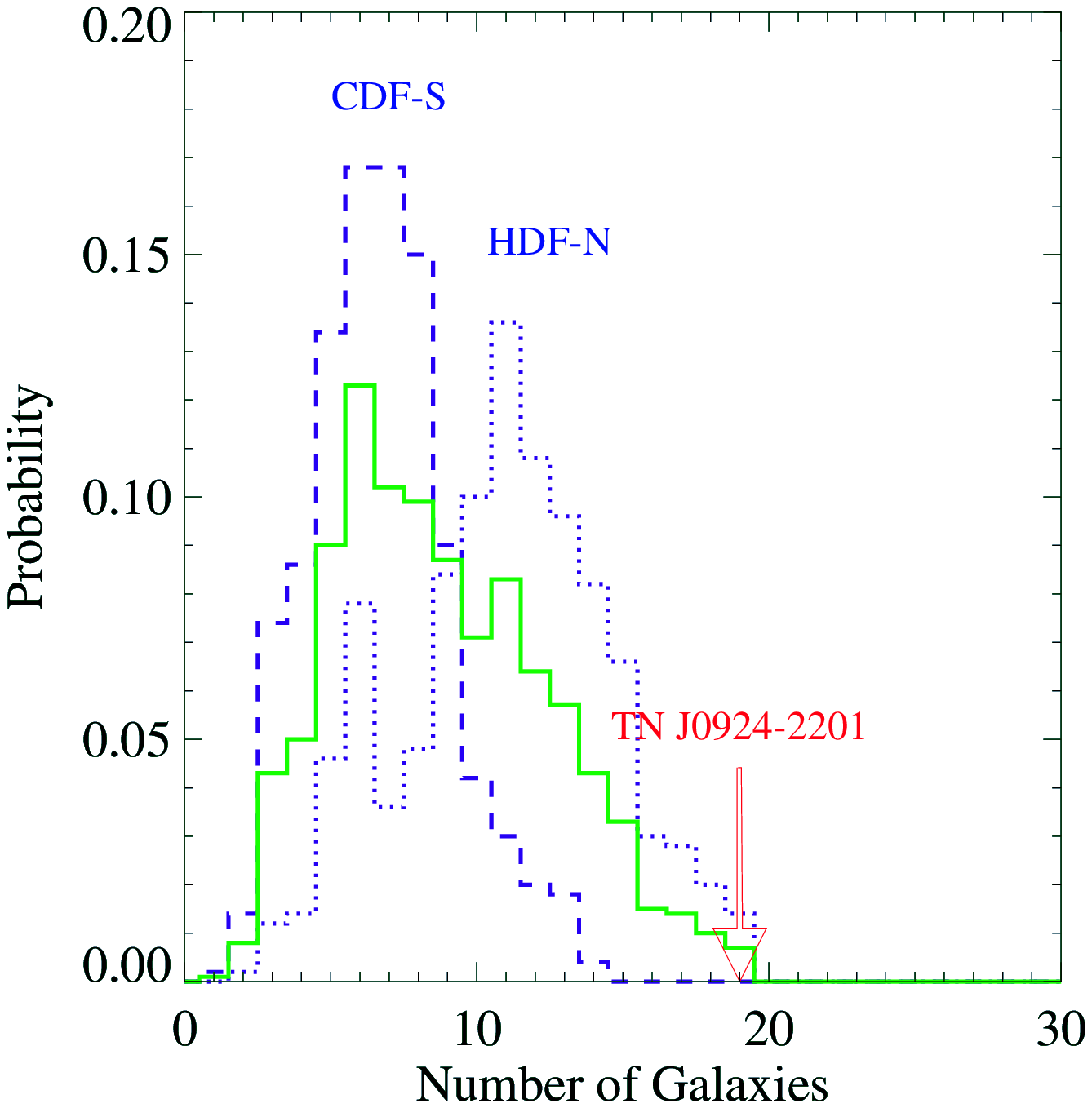}
\end{center}
\caption{\label{fig:cells}Histogram of counts-in-cells for the GOODS fields. The number of \vp-break objects were 
counted in $500$ randomly placed, square cells of 11.7 arcmin$^{2}$ in both the CDF-S (dashed line) and the HDF-N (dotted line). 
 The sum of the GOODS histograms is indicated (solid line). The total probability  
that one finds 19 objects in a single pointing in GOODS amounts to $\sim1$\%. The number of \vp-break objects 
detected in the TN J0924--2201 field (red arrow) is anomalously high compared to GOODS. This is evidence for a population of 
\vp-dropouts associated with the radio galaxy and the \lya\ emitters of \citet{venemans04}.}
\end{figure}

In this section we will test whether the overdensity of \lya\ emitters near TN J0924--2201 found by
 \citet{venemans05} is accompanied by an overdensity of \vp-dropout galaxies. 
To establish what the surface density is of \vp-dropouts in the `field' we have applied our selection criteria 
to the GOODS and the UDF Parallel fields. The color-color diagram for the objects in the GOODS fields 
is shown in Fig. \ref{fig:cccontrol}. The GOODS and UDF Parallel fields cover a total area of $\sim337$ 
arcmin$^{2}$ compared to $\sim12$ arcmin$^{2}$ for TN J0924--2201. At \zp$<26.5$ the numbers of \vp-dropouts 
satisfying our selection criteria are 277 for the combined GOODS fields, and 8 and 16 objects in the UDF-P1 
and UDF-P2 fields, respectively. Similar to \citet{bouwens05_goods} we find that the number of \vp-dropouts 
is ~10\% higher in the HDF-N compared to the CDF-S due to cosmic variance. We derive an average surface 
density of 0.9 arcmin$^{-2}$ 
for the field, consistent with \citet{giavalisco04_results} and the publicly available GOODS V1.1 catalogues which 
we have used to cross-check our results. The \vp-dropout surface density of 2.0 arcmin$^{-2}$ in the 
TN J0924--2201 field is twice as high, while the overall object surface densities at \zp$<26.5$, $S/N>5$ 
and $S/G<0.85$ are fairly constant: 106 arcmin$^{-2}$, 119 arcmin$^{-2}$ and 98 arcmin$^{-2}$ for the 
GOODS CDF-S, HDF-N, and TN J0924--2201 fields, respectively.

What is the significance of this factor 2 surface overdensity? LBGs are known to be strongly 
clustered at every redshift \citep{porciani02,ouchi04_r0}, and are known to have large field-to-field variations. 
In our particular case, it is interesting to calculate the probability of finding a certain number of \vp-dropouts 
in a single ACS pointing. Here we will use the additional constraint on the \ip--\zp\ color specified in 
Eq. \ref{eq:criteria2} and which is indicated by the shaded areas in Figs. \ref{fig:cc} and \ref{fig:cccontrol}. 
The angular distributions of the 218 GOODS \vp-dropouts satisfying these criteria are shown in 
Fig. \ref{fig:2dgoods}. The distribution appears filamentary with noticable `voids' that are 
somewhat smaller than one ACS pointing. To the lower-left of the GOODS HDF-N mosaic in Fig. \ref{fig:2dgoods} 
we have indicated the size of a single $3\farcm4\times3\farcm4$ ACS pointing for comparison. 
We measured the number of LBGs in 1000 (500 for each GOODS field) square 11.7 arcmin$^{2}$ cells placed at 
random positions and orientation angles. The cells were allowed to overlap, and are therefore not totally independent. 
The histogram of counts-in-cells is shown in Fig. \ref{fig:cells}. 
The number of LBG candidates in TN J0924--2201 falls on the extreme right 
of the expected distribution based on GOODS (indicated by the red arrow). None of the 
cells randomly drawn from the CDF-S contained 19 objects (the highest being 14), while the chance of 
finding 19 objects in a single pointing in the HDF-N was slightly over 1\%. Combining these results, 
TN J0924--2201 is overdense at the $>99$\% level with respect to GOODS. 

\begin{figure}[t]
\begin{center}
\includegraphics[width=0.5\textwidth]{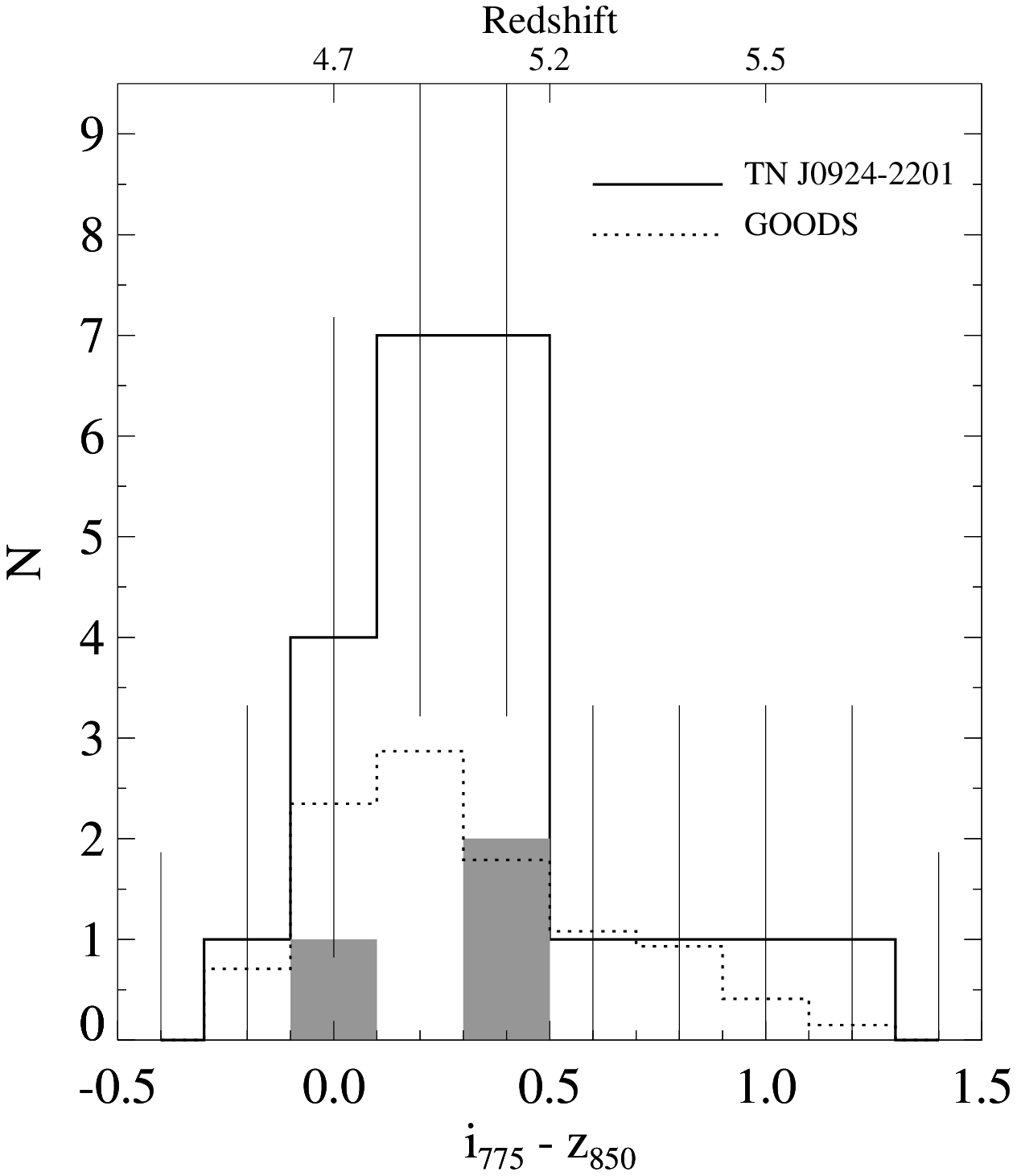}
\end{center}
\caption{\label{fig:iz}\ip--\zp\ color distribution of the TN J0924--2201 sample (solid curve). The GOODS color 
distribution, normalised to the area of the TN J0924--2201 field, is shown for comparison (dotted curve). 
The error bars are Poissonian in the low counts regime \citep{gehrels86}. The \ip--\zp\ color of the radio 
galaxy and the two \lya\ emitters that are included in the TN J0924--2201 sample are indicated by the shaded regions. 
The overdensity in the TN J0924--2201 field is most prominent at $0.0<$\ip--\zp$<0.5$. A slight excess in the number counts is seen 
around \ip--\zp$\approx0.5$, which corresponds to $z\approx5.2$ (see top axis) for typical values of $\beta$ 
(i.e., --1.5 to --2). 
}
\end{figure}

\begin{figure}[t]
\begin{center}
\includegraphics[width=0.5\textwidth]{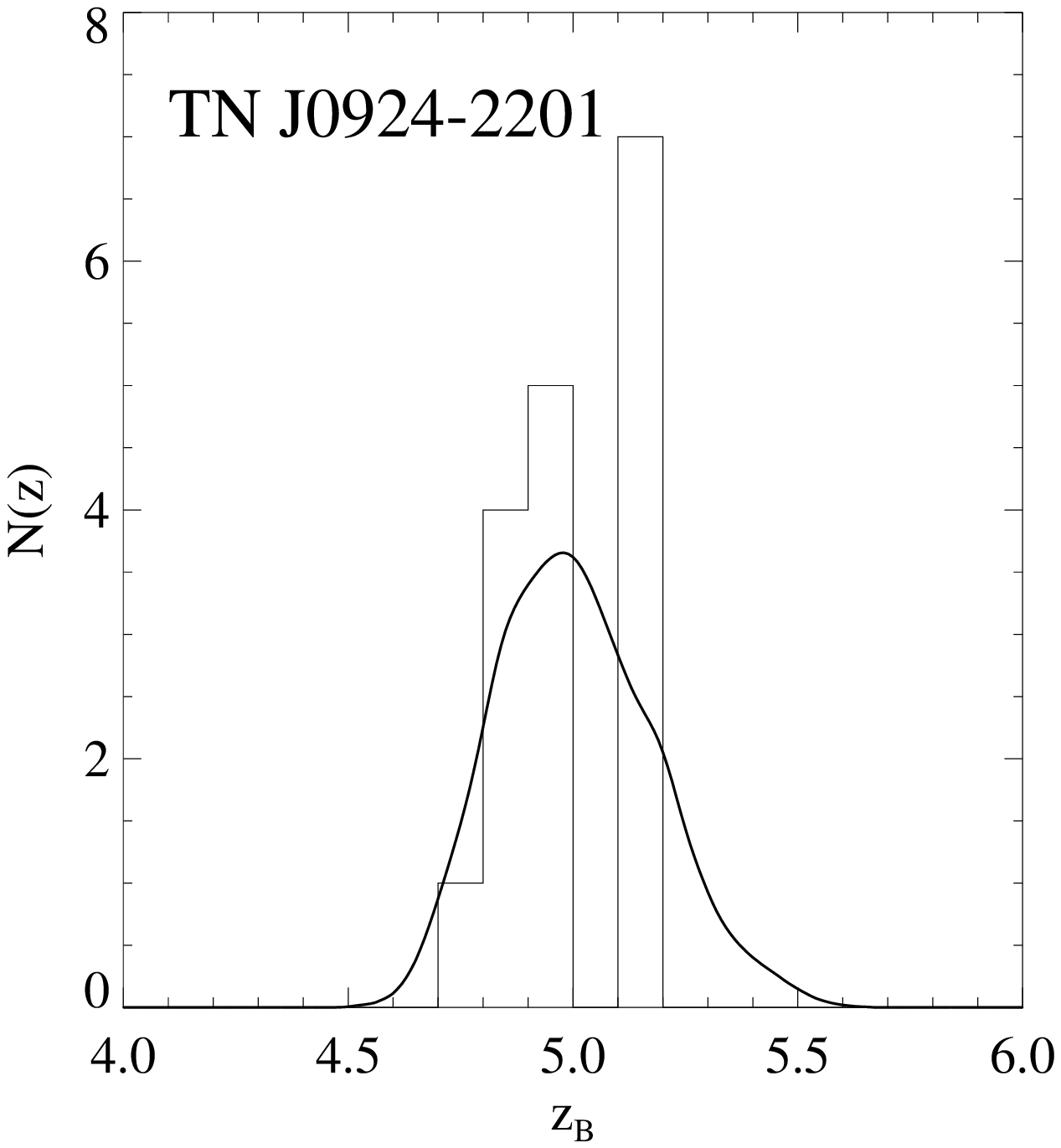}
\end{center}
\caption{\label{fig:bpz}Bayesian photometric redshift histogram for \vp-dropouts in TN J0924--2201 having $0.0<$\ip--\zp$<1.0$ and ODDS$>0.95$. BPZ was used with the standard redshift prior that is based on the magnitudes of galaxies in 
the HDF-N \citep[see][]{benitez04}. The total $z_B$ probability distribution (thick solid curve) has also been indicated.}
\end{figure}

As shown in Fig. \ref{fig:iz}, the excess in the TN J0924--2201 field over that of the GOODS sample 
(normalised to the same area) is primarily due to objects having \ip--\zp$\sim0.0-0.5$. This clustering 
observed in the \ip--\zp\ color distribution suggests that the significance of the surface overdensity 
is in fact much higher than the $>99$\% estimated above, given that $\sim30$\% of the GOODS 
\vp-dropouts populate the color diagram at \ip--\zp$>0.5$, compared to only $\sim10$\% of the TN J0924--2201 candidates.
The most significant number excess manifests itself around \ip--\zp$\approx0.5$, which matches the expected 
color of an LBG spectrum at the redshift of the radio galaxy assuming a typical slope of $\beta\approx-2$ 
(the approximate redshift for such a template spectrum is indicated on the top axis of Fig. \ref{fig:iz}). 
Two of the three previously known protocluster members that are in the \vp-dropout sample also lie near this color
(shaded regions in Fig. \ref{fig:iz}). Estimates of the photometric redshifts with BPZ also show a preference for $z\approx5.2$ and slightly lower redshifts, although the errors on $z_B$ are quite large ($\sim0.7$, Fig. \ref{fig:bpz} \& Table \ref{tab:lbgs}).
The subclustering in \ip--\zp\ and the photometric redshifts provide further 
evidence that the overdensity is associated with the radio galaxy and the \lya\ emitters. 

\begin{figure}[t]
\begin{center}
\includegraphics[width=0.5\textwidth]{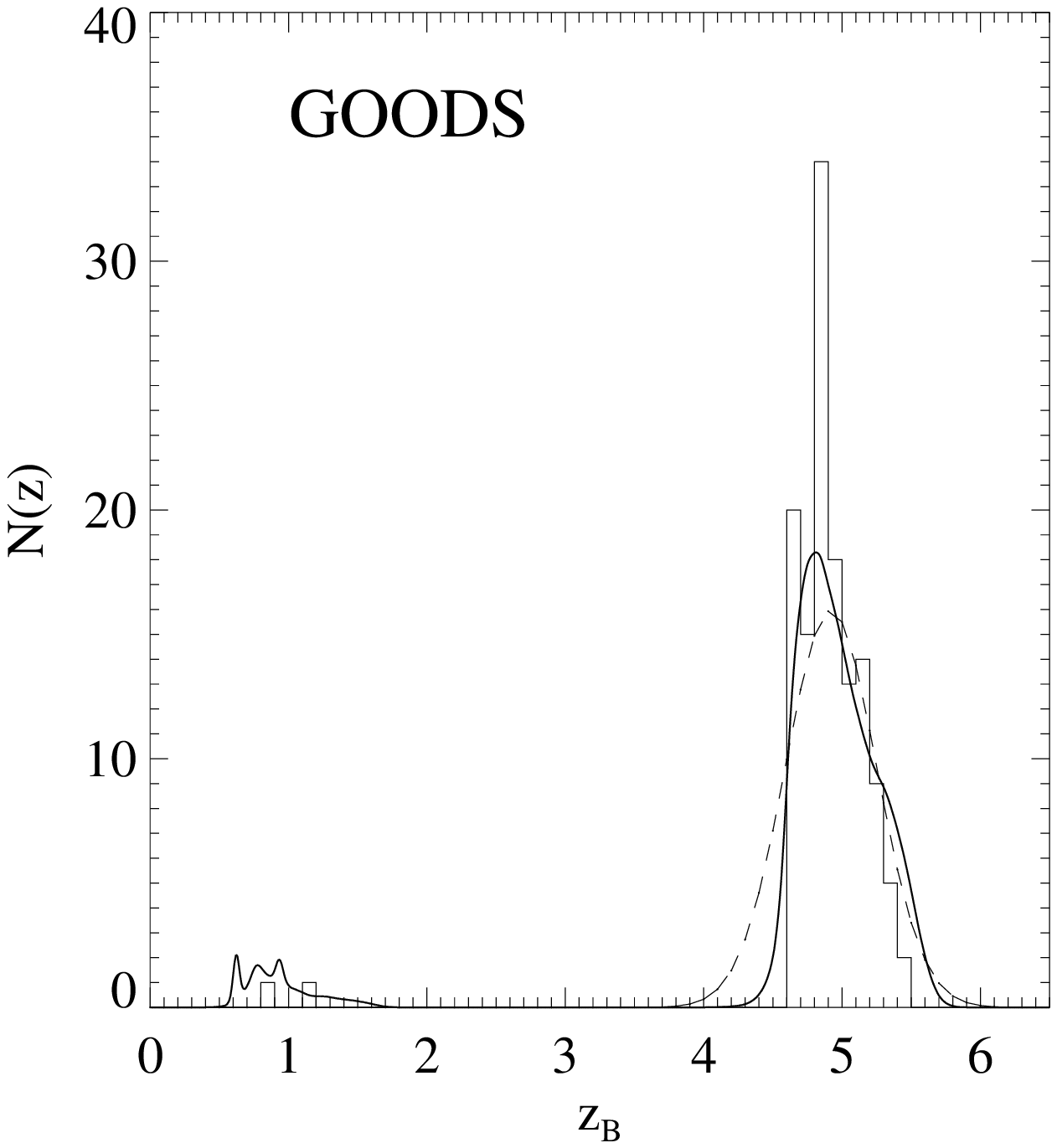}
\end{center}
\caption{\label{fig:goodsbpz}Photometric redshift histogram for the GOODS sample. Although the sample was 
selected using only the \vp\ip\zp\ passbands, redshifts were calculated using the full \bp\vp\ip\zp\ catalogue. 
The total $z_B$ probability distribution (thick solid curve) suggests a low redshift contamination of $\sim9\%$. 
The redshift distribution of \citet{giavalisco04_results} has been indicated for comparison (dashed curve). 
Our distribution is narrower because of the additional constraints on \ip--\zp\ (Eq. \ref{eq:criteria2}).}
\end{figure}

We can derive the \vp-dropout number densities from the comoving volume occupied by the objects. 
For the comoving volume one usually defines an effective volume, $V_{eff}$, that takes into account 
the magnitude and color incompletenesses. We estimated the effective redshift distribution, $N(z)$, associated 
with our selection criteria by running BPZ on the \bp\vp\ip\zp\ photometry of the large GOODS \vp-dropouts 
sample. The redshift distribution is shown in Fig. \ref{fig:goodsbpz}, where we have also indicated the 
sum of the redshift probability curves of each object to maintain information on secondary maxima, as well 
as the uncertainties associated with each object. Our effective redshift distribution is slightly 
narrower than the redshift distribution of \citet{giavalisco04_results} (indicated by the dashed line in Fig. 
\ref{fig:goodsbpz}), due to our additional constraint on \ip--\zp. Because we only used objects for which 
$z_B$ was relatively secure (i.e., objects having ODDS$>0.95$), as well as using the full $z_B$ probability 
curves to construct Fig. \ref{fig:goodsbpz}, we believe that our $N(z)$ is a good approximation to the true 
underlying redshift distribution. While our $N(z)$ could appear too narrow if the errors on $z_B$ are 
significantly underestimated, we can expect it to be narrower than that of \citet{giavalisco04_results} in any case.
The total probability of contamination seen around $z\sim1$ amounts to $\sim10$\%. This is similar to the number of objects in the GOODS sample for which the $S/N$ in \bp\ is $>2$. 

Using the effective $N(z)$ the comoving volume for the combined GOODS fields becomes $\sim5.5\times10^5$ Mpc$^3$. 
Here it is assumed that the selection efficiency at the peak of the redshift distribution is close 
to unity. Taking into account an incompleteness of $\sim50$\% for \zp$<26.5$  
\citep[from][]{giavalisco04_results} gives an effective volume twice as small and a GOODS \vp-dropout 
volume density of $8\times10^{-4}$ Mpc$^{-3}$. 
For TN J0924--2201, the effective volume is $\sim1\times10^4$ Mpc$^{3}$ giving a number density 
of $2\times10^{-3}$ Mpc$^{-3}$ if all galaxies are spread out across the volume. If, on the other hand, 
a significant fraction (e.g., $\gtrsim$50\%)
of the objects are associated with the radio galaxy and \lya\ emitters (assuming an effective protocluster volume 
of $8\times10^2$ Mpc$^{3}$ at $\bar{z}=5.2$ with $\Delta z=0.03$), 
we find a volume density of $\gtrsim1\times10^{-2}$ Mpc$^{-3}$ and a SFR density of $\gtrsim1\times10^{-1}$ M$_\odot$ yr$^{-1}$ Mpc$^{-3}$. This is at least a tenfold increase compared to that of the field.

One may wonder what the cosmic variance implies for a field as large as GOODS. \citet{somerville04} have presented 
a useful recipe for deriving the cosmic variance based on the clustering of dark matter halos in the analytic CDM model of \citet{sheth99}. 
Once the number density and mean redshift of a given population are known, one can derive the bias parameter, $b$, and calculate the 
variance of the galaxy sample, $\sigma_g=b\sigma_{DM}$, where $\sigma_{DM}$ is the variance of the 
dark matter. A number density of $\sim1\times10^{-3}$ Mpc$^{-3}$ corresponds to $b\approx4$ and a  
variance $\sigma_{DM}\approx0.07$ for dark matter haloes at $z\sim5$. This would imply that the upper limit 
for the cosmic variance of 
\vp-dropouts in a field as large as one of the GOODS fields is $\sim30$\%. The difference in the object densities 
that we found was $\sim10$\% between the two GOODS fields. Assuming that the CDF-S represents the absolute minimum 
of the allowed range would imply that new fields may be discovered showing significantly more sub-clustering 
on the scale of a single ACS pointing than currently observed. In the other extreme case that the HDF-N represents 
the absolute maximum, the TN J0924--2201 field should exhibit one of the highest surface densities of \vp-dropouts 
expected.   

The surface overdensity of \lya\ emitters around TN J0924--2201 was 1.5--6 compared to the field \citep{venemans04}. 
Our results would be marginally consistent with the lower value of $\sim2$. However, only two of the \lya\ emitters are bright 
enough to be included in our LBG sample. If the fraction of LBGs with high rest-frame equivalent width \lya\ in protoclusters is similar 
to that of the field (\citet{shapley03} find $\sim25$\%),  
$\sim6$ additional (i.e., non-\lya) `protocluster' LBGs are expected among our sample of 16 candidates (19 when including the radio galaxy and the two \lya\ emitters). Such an overdensity could 
easily be accommodated given the relative richness of LBGs in this field, although its ultimate verification must await spectroscopic follow-up. 

Based on the clustering statistics of relatively bright ($z^\prime<25.8$) $Vi^\prime z^\prime$-selected LBGs 
at $z\sim5$, \citet{ouchi04_r0} found that these objects are likely to be hosted by very massive dark matter 
halos of $\sim10^{12}$ M$_\odot$. 
The halo occupation number for these LBGs is almost unity, implying that almost every halo of this mass 
is expected to host a UV-bright LBG. Our sample contains several \vp-dropouts which have \zp$<25.0$ (the 
brightest being \#1873 with \zp$=24.2$), implying present-day halo masses of $\langle M(z=0)\rangle>10^{14}$ M$_\odot$. Whether any of these objects are associated with the radio galaxy should be confirmed by spectroscopy. 

\subsection{The host galaxy of TN J0924--2201}

The high radio luminosity of TN J0924--2201 indicates that it hosts a supermassive black hole, 
which must have acquired its mass in less than $\sim1$ Gyr.  
However, in many other respects we found that it appears unremarkable when compared to general Lyman break galaxies at a similar redshift. 
Although there is a wide dispersion in the properties of the 
highest redshift radio galaxies \citep[e.g.][]{rawlings96,dey97,reuland03,zirm05} 
it might be interesting to naively compare TN J0924--2201 to TN J1338--1942 at $z=4.1$ also studied with ACS \citep{zirm05}.  
The optical host of TN J0924--2201 is almost 2 magnitudes fainter 
(at similar rest-frame wavelengths) than TN J1338--1941. There are several \vp-dropouts in our sample that have 
brighter magnitudes (and therefore higher SFRs) than the radio source, while TN J1338--1942 is by far the brightest object among 
the sample of associated \gp-dropouts found in that field \citep{miley04}. Likewise, TN J0924--2201 has a size that is comparable to the average size of \vp-dropouts \citep{bouwens04,ferguson04}, while TN J1338--1942 is an exceptionally large ($\sim2$\arcsec) galaxy. 
If TN J0924--2201 is to develop into a similar source within the $\sim400$ Myr or so between 
$z\sim5$ and $z\sim4$, it would require an increase in the projected radio source size by a factor $\sim 5$, 
in \lya\ luminosity by a factor $\sim 60$, in SFR  
by a factor of $\sim10$, and in UV size by at least a factor of 2. The recent detection of molecular gas (CO) by \citet{klamer05} suggests that there is $\sim10^{11}M_\odot$ of (inferred) gas mass present. 
The rapid enrichment that brought about this reservoir of molecular 
gas could have been facilitated by the early formation of the radio source and the triggering of 
massive star formation. The amount of gas present shows that there is plenty of material available to sustain 
a high SFR of several $100M_\odot$ yr$^{-1}$, possibly allowing this source to undergo dramatic changes in its 
UV luminosity and morphology during certain stages of its evolution.

\section{Conclusions}

We have presented statistical evidence for an overdensity of star forming galaxies associated with the radio galaxy TN J0924--2201. 
Our result is consistent with the overdensity of \lya\ emitters discovered previously by 
\citet{venemans04}, and is comparable to overdensities of \lya\ emitters and Lyman break galaxies found around other 
high redshift radio galaxies. TN J0924--2201 could be a protocluster that will evolve into a cluster with a mass of $\sim10^{14}$ M$_\odot$ at $z=0$. 

The existence of relatively massive structures in the early Universe may not be uncommon, as suggested by, for example, 
the existence of quasars at even higher redshifts \citep[e.g.][]{fan03}, the evidence for protoclusters out to $z<6$,  
and the increasingly higher limit that can be set 
on the redshift of reionization. Regions of high mass concentrations are rare, strongly clustered objects 
at every redshift, that underwent high amplification since the initial conditions \citep{kaiser84}. 
Radio galaxies 
are suspected to be the 
sites of the formation of a massive galaxy. The evidence reported of in this paper contributes to the hypothesis 
that redshift filaments and possibly groups or clusters of galaxies emerged together with these massive galaxies. In the radio-loud AGN 
unification model, the viewing 
angle relative to the jet determines whether a galaxy will be seen as a radio galaxy or as a radio-loud quasar \citep{barthel89}. It is 
therefore expected that early galaxy overdensities could, in principle, also be found around high redshift radio-loud quasars.  
Results indicate that the same may hold for at least some radio-quiet quasars at $z>5$ too \citep[][Stiavelli et al. 2005]{djorgovski99,djorgovski03}. 

Although primordial galaxy overdensities so far discovered are not solely limited to fields that contain a luminous AGN  \citep[e.g.][]{steidel99,ouchi05}, they hold a strong connection to low redshift clusters due to the presence 
of a supermassive black hole and possibly a developing brightest cluster galaxy \citep{zirm05}. When the radio 
source has switched off, the host galaxy may become indistinguishable from other relatively massive, quiescent 
galaxies that may have been active in the past. The clustering properties and inferred halo masses of the brightest 
Lyman break galaxies suggest that their descendants at $z=0$ fall into groups and clusters of galaxies \citep{ouchi04_r0}. 
Interestingly, the number densities of protoclusters, radio galaxies and Lyman break galaxy redshift spikes have 
all been found to be consistent with the abundance of local clusters \citep[e.g.][and references therein]{steidel99,venemans02,ouchi05}. Structures like TN J0924--2201 provide ideal comparisons with state-of-the-art $N$-body simulations. 
\citet{springel05} found that the descendants of the most massive objects at high redshift (presumed to be luminous quasars) can almost 
exclusively be identified with the most massive clusters at the current epoch.
It might become possible to trace back cluster evolution from the well-studied regimes at $z\lesssim1$ to slight overdensities at very high redshifts, possibly up to the epoch of reionization.

\begin{acknowledgements}
RAO is very grateful to Masami Ouchi for invaluable discussions, to 
Mike Dopita for discussion of Fig. \ref{fig:sb99}, and to Carlos De Breuck for 
providing the radio map of TN J0924--2201 shown in Fig. \ref{fig:rg}. 
We thank the anonymous referee for helping to improve the manuscript.

ACS was developed under NASA contract NAS 5-32865, and this research
has been supported by NASA grant NAG5-7697 and
by an equipment grant from Sun Microsystems, Inc.
The {Space Telescope Science Institute} is operated by AURA Inc., under
NASA contract NAS5-26555. We are grateful to K.~Anderson, J.~McCann,
S.~Busching, A.~Framarini, S.~Barkhouser, and T.~Allen for their
invaluable contributions to the ACS project at JHU.
\end{acknowledgements}





\begin{deluxetable}{lcrrlc} 
\tablecolumns{6} 
\tablewidth{0pc} 
\tablecaption{\label{tab:log2}Observational Data} 
\tablehead{\multicolumn{1}{c}{Field} &\multicolumn{1}{c}{Filter} & \multicolumn{1}{c}{$T_{exp}$} & \multicolumn{1}{c}{$A$} & \multicolumn{1}{c}{Zeropoint}& \multicolumn{1}{c}{Area}\\
 \colhead{}  & \colhead{}                 & \colhead{(s)}     &  \multicolumn{1}{c}{(mag)}             &  \multicolumn{1}{c}{(mag)}& \multicolumn{1}{c}{(arcmin$^2$)}}
\startdata 
TN J0924--2201 & \vp\   &   9400& 0.167 & 36.4262$^a$& 11.7\\
TN J0924--2201 &\ip\   &  11800& 0.115 & 35.8202$^a$& 11.7\\
TN J0924--2201 &\zp\  &  11800& 0.080 & 35.0228$^a$& 11.7\\
\hline
GOODS CDF-S &\vp\  &   5120& 0.023 & 26.4934$^b$ & 156 \\
GOODS CDF-S &\ip\  &   5120& 0.017 & 25.6405$^b$  & 156  \\
GOODS CDF-S &\zp\  &  10520& 0.012 & 24.8432$^b$  & 156  \\
\hline
GOODS HDF-N &\vp\  &   5000& 0.035 & 26.4934$^b$  & 158  \\
GOODS HDF-N &\ip\  &   5000& 0.024 & 25.6405$^b$  & 158  \\
GOODS HDF-N &\zp\  &  10660& 0.018 & 24.8432$^b$  & 158  \\
\hline
UDF-PARALLEL 1 &\vp\  &   22300& 0.023 & 37.3571$^a$& 11.7 \\
UDF-PARALLEL 1 &\ip\  &   41400& 0.016 & 37.1970$^a$ & 11.7 \\
UDF-PARALLEL 1 &\zp\   &   59800& 0.012 & 36.8038$^a$& 11.7  \\
\hline
UDF-PARALLEL 2 &\vp\  &   20700& 0.026 & 37.2763$^a$& 11.7  \\
UDF-PARALLEL 2 &\ip\  &   41400& 0.018 & 37.1970$^a$ & 11.7 \\
UDF-PARALLEL 2 &\zp\   &   62100& 0.013 & 36.8448$^a$ & 11.7 \\
\enddata 
\tablenotetext{a}{Zeropoint for the total exposure \citep{sirianni05}.}
\tablenotetext{b}{Zeropoint for a 1 s exposure \citep{giavalisco04_survey}.}
\end{deluxetable}

\begin{deluxetable}{llllrrrlcc}
\tabletypesize\tiny
\tablecolumns{10} 
\tablewidth{0pc} 
\tablecaption{\label{tab:laes}Properties of the spectroscopically confirmed Ly$\alpha$\ emitters.}
\tablehead{\multicolumn{1}{c}{ID$^a$} & \multicolumn{1}{c}{$\alpha_{J2000}$} & \multicolumn{1}{c}{$\delta_{J2000}$} & \multicolumn{1}{c}{\vp--\ip$^b$} & \multicolumn{1}{c}{\ip--\zp$^b$} & \multicolumn{1}{c}{\zp$^c$} & \multicolumn{1}{c}{$z_{spec}^d$} & \multicolumn{1}{c}{$r_{hl,i}^e$} & \multicolumn{1}{c}{$r_{hl,z}^f$} & \multicolumn{1}{c}{SFR$^g$} }
\startdata 
RG$^h$   & 09:24:19.89 & --22:01:41.23 & $>2.73$       & $0.40\pm0.13$  & $25.45\pm0.12$ & $5.199$ & 0.19\arcsec & 0.23\arcsec & $13.8^{+2.2}_{-1.9}$\\ 
2881$^i$ & 09:24:23.87 & --22:03:43.97 & $2.48\pm0.29$ & $0.37\pm0.09$  & $25.80\pm0.09$ & $5.168$ & 0.09\arcsec & 0.11\arcsec & $9.7^{+0.6}_{-0.6}$\\ 
1388$^i$ & 09:24:16.66 & --22:01:16.41 & $2.20\pm0.36$ & $0.08\pm0.17$  & $26.33\pm0.17$ & $5.177$ & 0.12\arcsec & 0.15\arcsec & $5.9^{+0.8}_{-0.6}$\\
2849$^i$ & 09:24:24.29 & --22:02:30.11 & $>1.85$       & $-0.27\pm0.43$ & $27.06\pm0.25$ & $5.177$ & 0.13\arcsec & 0.16\arcsec & $3.0^{+0.8}_{-0.6}$\\
2688$^i$ & 09:24:25.65 & --22:03:00.27 & $1.16\pm0.42$ & $<-0.49$ & $>28.35$ & $5.173$ & 0.08\arcsec & -- & $<1$\\
\enddata
\tablenotetext{a}{IDs are from \citet{venemans05}.}
\tablenotetext{b}{Isophotal color, using $2\sigma$ limits in \vp.}
\tablenotetext{c}{Kron magnitude.}
\tablenotetext{d}{Spectroscopic redshifts from \citet{venemans05}.}
\tablenotetext{e}{Half-light radius measured in \ip.}
\tablenotetext{f}{Half-light radius measured in \zp.}
\tablenotetext{g}{SFR measured in \zp\ in M$_\odot$\ yr$^{-1}$.}
\tablenotetext{h}{Detection based on the \zp\ image.}
\tablenotetext{i}{Detection based on the \ip\ image.}
\end{deluxetable}

\begin{deluxetable}{llllrrrrr}
\tabletypesize\tiny
\tablecolumns{9} 
\tablewidth{0pc} 
\tablecaption{\label{tab:lbgs}Properties of \vp-dropouts in the field of TN J0924--2201.}
\tablehead{\multicolumn{1}{c}{ID} & \multicolumn{1}{c}{$\alpha_{J2000}$} & \multicolumn{1}{c}{$\delta_{J2000}$} & \multicolumn{1}{c}{\vp--\ip$^b$} & \multicolumn{1}{c}{\ip--\zp$^b$} & \multicolumn{1}{c}{\zp$^c$} & \multicolumn{1}{c}{$r_{hl,z}^d$} & \multicolumn{1}{c}{SFR$^e$} & \multicolumn{1}{c}{$z_B^f$}}
\startdata 
  $  1873$       &09:24:15.17 &     --22:01:53.2   &$2.05\pm0.09$&$       0.37\pm0.04$&$24.21\pm0.04$&$0\farcs18$& 42&$5.19^{+0.73}_{-0.73}$\\
  $   119$       &09:24:29.06 &     --22:02:41.1  &$1.79\pm0.12$&$        0.08\pm0.07$&$24.82\pm0.09$&$0\farcs24$& 24 &$4.74^{+0.68}_{-0.68}$\\
  $   303 $      &09:24:29.03 &     --22:01:53.2  &$2.04\pm0.13$&$        0.23\pm0.06$&$24.82\pm0.07$&$0\farcs22$& 24 &$4.86^{+0.69}_{-0.69}$\\
  $   444 $      &09:24:28.11 &     --22:01:47.4  &$2.35\pm0.32$&$        0.16\pm0.12$&$25.20\pm0.11$&$0\farcs28$& 17 &$4.93^{+0.70}_{-0.70}$\\
  $   1814 $     &09:24:19.78 &     --22:59:58.1  &$ >2.18$&$             1.07\pm0.15$&$25.29\pm0.09$&$0\farcs24$& 16 &$5.56^{+0.77}_{-0.77}$\\
  $   310 $      &09:24:28.10 &     --22:02:18.1  &$1.93\pm0.18 $&$       0.47\pm0.08$&$25.42\pm0.08$&$0\farcs11$& 14 &$5.23^{+0.73}_{-0.73}$\\
$1396$ (RG)$^a$  &09:24:19.91 &     --22:01:41.7   &$ >2.73   $&$         0.40\pm0.13$&$25.45\pm0.12$&$0\farcs23$& 13&$5.16^{+0.72}_{-0.72}$\\
$    1979$       &09:24:12.50 &     --22:02:45.5  &$2.09\pm0.19 $&$       0.09\pm0.09$&$25.46\pm0.12$&$0\farcs14$& 13 &$4.90^{+0.69}_{-0.69}$\\
$    1802 $      &09:24:14.92 &     --22:02:15.8  &$2.41\pm0.26$&$        0.29\pm0.09$&$25.46\pm0.10$&$0\farcs15$& 13 &$5.01^{+0.71}_{-0.71}$\\
$     871 $      &09:24:25.76 &     --22:01:11.4  &$2.03\pm0.41$&$        0.79\pm0.14$&$25.50\pm0.13$&$0\farcs23$& 13 &$5.39^{+0.75}_{-3.90}$\\
$     595 $      &09:24:27.00 &     --22:01:37.6  &$>2.55      $&$        0.48\pm0.14$&$25.53\pm0.14$&$0\farcs23$& 12 &$5.21^{+0.73}_{-0.73}$\\
$     894 $      &09:24:23.15 &     --22:02:16.6  &$2.27\pm0.20 $&$       0.15\pm0.08$&$25.53\pm0.09$&$0\farcs15$& 12 &$4.92^{+0.70}_{-0.70}$\\
$    1047  $     &09:24:22.19 &     --22:01:59.6  &$1.74\pm0.16 $&$     -0.08\pm0.11$ &$25.65\pm0.15$&$0\farcs20$& 11 &$4.76^{+0.68}_{-0.68}$\\
$449$ (2881)$^a$ &09:24:23.89 &     --22:03:44.4 &$2.48\pm0.29 $&$        0.37\pm0.09$&$25.80\pm0.09$&$0\farcs11$& 9.7  &$5.13^{+0.72}_{-0.72}$\\
$    1736  $     &09:24:18.92 &     --22:00:42.2  &$1.84\pm0.18 $&$       0.19\pm0.11$&$25.92\pm0.12$&$0\farcs11$& 8.6&$4.77^{+0.68}_{-0.68}$\\
$     670 $      &09:24:25.41 &     --22:02:07.2  &$2.16\pm0.35$&$        0.28\pm0.14$&$25.93\pm0.16$&$0\farcs17$& 8.5&$5.02^{+0.71}_{-0.71}$\\
$     739 $      &09:24:25.76 &     --22:01:43.0  &$2.84\pm0.50$&$        0.16\pm0.12$&$26.09\pm0.15$&$0\farcs12$& 7.4&$4.94^{+0.70}_{-0.70}$\\
$    1074 $      &09:24:21.24 &     --22:02:21.6  &$1.98\pm0.30$&$       -0.16\pm0.18$&$26.17\pm0.18$&$0\farcs18$& 6.9&$4.79^{+0.68}_{-0.68}$\\
$     510 $      &09:24:28.81 &     --22:01:12.2  &$>2.02      $&$        1.21\pm0.18$&$26.27\pm0.12$&$0\farcs09$& 6.2&$5.62^{+0.78}_{-0.78}$\\
$    1385 $      &09:24:19.31 &     --22:02:01.8  &$2.10\pm0.43 $&$       0.48\pm0.17$&$26.30\pm0.15$&$0\farcs15$& 6.1&$5.18^{+0.73}_{-0.73}$\\
$1844$ (1388)$^a$&09:24:16.68 &     --22:01:16.8  &$2.20\pm0.36 $&$       0.08\pm0.16$&$26.33\pm0.17$&$0\farcs15$& 5.9&$4.87^{+0.69}_{-0.69}$\\
$     505  $     &09:24:25.42 &     --22:02:48.0  &$>2.00      $&$        0.52\pm0.23$&$26.36\pm0.22$&$0\farcs16$& 5.8&$5.23^{+0.73}_{-1.18}$\\
$    1898  $     &09:24:14.17 &     --22:02:15.9  &$>2.41     $&$         0.44\pm0.17$&$26.49\pm0.19$&$0\farcs12$& 5.1&$5.18^{+0.73}_{-0.73}$\\
\enddata 
\tablenotetext{a}{IDs between parentheses refer to \citet{venemans05} and Table \ref{tab:laes}.}
\tablenotetext{b}{Isophotal color, using $2\sigma$ limits in \vp}
\tablenotetext{c}{Kron magnitude.}
\tablenotetext{d}{Half-light radius measured in \zp.}
\tablenotetext{e}{UV star formation rate measured in \zp\ in M$_\odot$\ yr$^{-1}$.}
\tablenotetext{f}{Bayesian photometric redshift.}
\end{deluxetable} 

\end{document}